\documentclass[aps,prd,nofootinbib,showkeys,preprint,floatfix]{revtex4}

\usepackage{amssymb}
\usepackage{amsmath}
\usepackage{amsfonts}
\usepackage{graphicx}
\usepackage{xcolor}
\usepackage{xspace}
\usepackage{ulem}
\usepackage{lscape}
\usepackage{wasysym }
\usepackage[toc,page]{appendix}
\usepackage{mathtools}
\usepackage{multirow,rotating}
\usepackage{stackrel}
\usepackage{cancel}
\usepackage{pifont}
\usepackage{dcolumn}% Align table columns on decimal point
\usepackage{bm}% bold math

\def\be{\begin{equation}}
\def\ee{\end{equation}}

\def\bee{\begin{eqnarray}}
\def\eee{\end{eqnarray}}
\def\im{\mathrm{i}}
\def\e{\mathrm{e}}
\def\beginm{\left(\begin{array}}
\def\endm{\end{array}\right)}

\newcommand{\eL}{\mathcal{L}}% Lagrangian
\newcommand{\V}{\mathcal{V}}% Potential
\newcommand{\D}{\ensuremath{\mathrm{D}}}% covariant derivative
\renewcommand{\d}{\ensuremath{\mathrm{d}}}% differential
\newcommand{\hc}{\ensuremath{\mathrm{h.c.}}}% hermitean conjugated part
\newcommand{\C}{\ensuremath{c}}% charge conjugation
\newcommand{\T}{\ensuremath{\mathrm{T}}}% matrix transposition
\newcommand{\SU}[1]{\ensuremath{\mathrm{SU}(#1)}}

\newcommand{\U}[1]{\ensuremath{\mathrm{U}(#1)}}

\newcommand{\sles}[1]{#1 \hspace{-6.5 pt}/}
\renewcommand{\d}{\ensuremath{\mathrm{d}}}% differential
\newcommand{\keV}{\ensuremath{\mathrm{\,keV}}}
\newcommand{\MeV}{\ensuremath{\mathrm{\,MeV}}}
\newcommand{\GeV}{\ensuremath{\mathrm{\,GeV}}}
\newcommand{\TeV}{\ensuremath{\mathrm{\,TeV}}}
\newcommand{\eV}{\ensuremath{\mathrm{\,eV}}}
\newcommand{\meV}{\ensuremath{\mathrm{\,meV}}}

\definecolor{Adam}{rgb}{0,0.5,0}
\definecolor{Sergey}{rgb}{0,0,1}
\newcommand{\Sergey}[1]{{\color{black}#1}}
\newcommand{\Adam}[1]{{\color{black}#1}}

\newcommand{\AddrUFSM}{
Universidad T\'ecnica Federico Santa Mar\'\i a, \\
Centro-Cient\'\i fico-Tecnol\'{o}gico de Valpara\'\i so, \\
Casilla 110-V, Valpara\'\i so,  Chile}
\newcommand{\AddrIEAPCTU}{Institute of Experimental and Applied Physics, Czech Technical University in Prague, Prague, Czech Republic}

\begin{document}

%\preprint{APS/123-QED}

\title{Low-scale seesaw from neutrino condensation
%\\ Observable seesaw from neutrino condensation\\ Low energy scale seesaw mechanism from neutrino condensation
}% Force line breaks with \\
\author{Claudio Dib}\email{claudio.dib@usm.cl}\affiliation{\AddrUFSM}
\author{Sergey Kovalenko}\email{sergey.kovalenko@usm.cl}\affiliation{\AddrUFSM}
\author{Ivan Schmidt}\email{ivan.schmidt@usm.cl}\affiliation{\AddrUFSM}
\author{Adam Smetana}\email{adam.ic.smetana@gmail.com}\affiliation{\AddrIEAPCTU}
%\author{Sergey Kovalenko, Ivan Schmidt, Adam Smetana}

\date{\today}



\begin{abstract}
Knowledge of the mechanism of neutrino mass generation would help understand a lot more about Lepton Number Violation (LNV), the cosmological evolution of the Universe, or the evolution of astronomical objects.
%
%Knowing how the mass generation of neutrinos is realized would help us to understand a lot more about the Lepton Number Violation (LNV), the cosmological evolution of the Universe, or the evolution of astronomical objects.
Here we propose a verifiable and viable extension of the Standard model for neutrino mass generation, with a low-scale seesaw mechanism via LNV condensation in the sector of sterile neutrinos. To prove the concept, we analyze a simplified model of just one single family of elementary particles and check it against a set of phenomenological constraints coming from electroweak symmetry breaking, neutrino masses, leptogenesis and dark matter. The model predicts (i) TeV scale quasi-degenerate heavy sterile neutrinos, suitable for leptogenesis with resonant enhancement of the $CP$ asymmetry, (ii) a set of additional heavy Higgs bosons whose existence can be challenged at the LHC, (iii) an  additional light and sterile Higgs scalar which is a candidate for decaying warm dark matter, and (iv) a majoron. Since the model is based on simple and robust principles of dynamical mass generation,
its parameters are very restricted, but remarkably it is still within current phenomenological limits.
%it does not allow to play too much with its parameters in order to escape the phenomenological limits. Therefore we consider the fact that the model has not been ruled out yet quite remarkable.
\end{abstract}

\maketitle

\section{Introduction}
\label{sec:Introduction}

The Lagrangian of the Standard model (SM) of elementary particles has an accidental $\U{1}_L$ symmetry of conservation of lepton number $L$.
%When one includes nonperturbative effects of axial anomalies, lepton number as part of the $\U{1}_{B+L}$ group is broken by a tiny amount,
%which is an effect
%a feature that could have
%been observed
%shown up only in the early Universe as transitions enhanced by sphalerons, allowing for the creation of baryon abundance via leptogenesis. Nowadays, such $(B+L)$-anomalous lepton number violation is practically unobservable.
%{\color{red} Therefore, lepton number remains as a well conserved quantum number, even after the electroweak symmetry breaking, connected with the generation of Dirac masses of fermions. Why should we then question lepton number conservation?
%\\ THIS IS NOT TRUE: L COULD BE BROKEN, EVEN THOUGH B+L BREAKING PROCESSES NO LONGER OCCUR.
%}
%
Nowadays there are at least three reasons why a sizable, beyond the SM, lepton number violation (LNV) should be considered. First, having a number of drawbacks such as the vacuum stability problem, lack of naturalness, or several hierarchy problems, the SM is increasingly understood as a low-energy effective model. As such, its renormalizable operators are just the leading-order terms in an infinite expansion of the effective Lagrangian, while the rest of the expansion consists of non-renormalizable operators of dimension larger than 4, which are suppressed by inverse powers of the scale of new physics. The least suppressed non-renormalizable operator, respecting all the SM local symmetries, is the dimension five Weinberg operator, which violates lepton number conservation by two units. Second, in order to explain naturally the smallness of neutrino masses, various seesaw mechanisms have been proposed, which rely on a LNV mixing of neutrino fields, giving rise to Majorana neutrino mass eigenstates. Third, LNV is a necessary condition for successful leptogenesis, which in turn could explain the baryon abundance of the Universe.

Usually these three aspects of LNV are jointly realized within extensions of the SM by seesaw mechanisms of various types, which provide neutrino masses naturally small compared to the charged fermion masses by means of a suppression coming from an inverse power of a large seesaw mass scale. The various types of seesaw mechanisms differ by the assumed origin of the seesaw mass scale as the mass of some new heavy fields, such as right-handed neutrinos, triplet scalar bosons, etc. Moreover, the size of the seesaw scale is not fixed purely by the size of the active neutrino masses, because other parameters may enter the neutrino mass formula. As such, the seesaw scale can have any value between the electroweak and Planck scales. Apart from theoretical restrictions in the form of postulating some symmetry or requiring some degree of naturalness, leptogenesis is what brings the most serious hints about the size of the seesaw scale, interpreted as a mass of heavy sterile Majorana neutrinos. This allows us to distinguish between high-scale and low-scale seesaw mechanisms. In fact, for successful leptogenesis, masses of the sterile neutrinos should be either very large, $10^{8}\GeV$ or more \cite{DAVIDSON200225}, or if smaller, then they should be quasi-degenerate in order to resonantly enhance the $CP$ asymmetry \cite{Pilaftsis:1997jf}, so defining the low-scale seesaw mechanism.

The low-scale seesaw mechanisms are attractive for their ability to offer an interesting phenomenology in the ballpark of current accelerator facilities. The extreme case, with quasi-degenerate sterile neutrinos of masses $\sim\mathcal{O}(10\GeV)$, is the phenomenologically successful $\nu$MSM model, based on a type-I seesaw mechanism \cite{Shaposhnikov2005}, which however requires tuning of a quite large number of free parameters, without any leading principle apart from phenomenological constraints. The linear and inverse seesaw mechanisms, on the other hand, contain heavy sterile Majorana neutrinos with quasi-degenerate masses in a more natural way, for the price of doubling the number of right-handed neutrino fields compared to the type-I seesaw mechanism. They also allow for setting up the seesaw scale to be lepton number conserving, which opens up the possibility of studying spontaneous LNV as a low-energy phenomenon. Motivated by these attractive features we elaborate our model on the basis of a combined inverse and linear low-scale seesaw mechanisms, where leptogenesis will provide the key ingredients to fix the model parameters.

The combined case of linear and inverse seesaw mechanisms is a natural consequence of the presence of two types of right-handed neutrinos \cite{Malinsky:2005bi}.
Various models, implementing this scenario, have been proposed in the literature (for a recent review, see, for instance, Ref.~\cite{Gavela:2009cd}).
%{and as such it has been already proposed several times in the literature starting from
%\cite{Malinsky:2005bi}.
%
Typically in these models the seesaw values of relevant parameters
are
%have been however
set by hand, either directly as a new mass parameter or indirectly as a free parameter of a corresponding Yukawa coupling.
%
%In these cases the seesaw values of relevant parameters have been however set {\it by hand}, either directly as a new mass parameter or indirectly as a free parameter of a corresponding Yukawa coupling.
%
In the present paper we propose  a {\it dynamical origin} of the neutrino mass parameters rooted in neutrino condensation.
%
%We want to avoid this in our model, wherefore we propose a {\it dynamical origin} of selected neutrino mass parameters, by means of neutrino condensation.
%
The idea is that due to some new attractive force felt by neutrinos, LNV vacuum neutrino condensates are formed, and meson-like new (pseodo-)scalar bosons emerge as composite states of the neutrino fields. The seesaw mass matrix elements are generated dynamically as the vacuum expectation values (VEVs) of the composite scalars.
%:{SK_11.04.2019-1} 
%The idea of neutrino condensation, which, however, 
%\textcolor{blue}{violates lepton number explicitly}, 
Neutrino mass models with neutrino condensation and 
explicit LNV, has already been studied in the literature \cite{Martin:1991xw, Antusch:2002xh, Smetana:2013hm,Smetana:2013exa}
%:{SK_11.04.2019-1} END
%
%The idea of neutrino condensation, which, however, conserves lepton number,  has already been studied in the literature \cite{Martin:1991xw, Antusch:2002xh, Smetana:2013hm,Smetana:2013exa}
in the context of  type-I seesaw mechanism. Here we apply a similar strategy, in the framework of the linear and inverse seesaw mechanisms, assigning lepton number to right-handed neutrinos and selecting their LNV condensation channels in such a way that the new composite scalars also carry lepton number.
%
%Such neutrino condensation in the framework of a type-I seesaw mechanism that has already been studied in the literature \cite{Martin:1991xw, Antusch:2002xh, Smetana:2013hm,Smetana:2013exa}. Here we apply the same strategy, in the framework of the linear and inverse seesaw mechanisms, assigning lepton number to right-handed neutrinos and selecting their condensation channels in such a way that the new composite scalars also carry lepton number.
%
The model Lagrangian is manifestly lepton number invariant and provides only lepton-number-conserving elements in the neutrino mass matrix. The lepton-number-violating elements, which trigger the combined linear and inverse seesaw mechanism, are dynamically generated, being proportional to VEVs of the composite scalars. Lepton number gets spontaneously broken and a massless composite majoron appears in the spectrum of the observable particles, along with a handful of other additional Higgs bosons.

In order to prove the phenomenological feasibility of our LNV neutrino condensation
%concept,
setup
we parametrize the new neutrino-attracting force by a simple-minded four-neutrino interaction. Such an approximation allows for an analysis of the low-energy particle spectrum to a sufficient detail by standard tools of renormalizable effective Lagrangians \cite{Bardeen:1989ds}. Even though there might be many non-perturbative aspects of the neutrino condensation inaccessible within this approach, we believe that the main qualitative features and quantitative estimates can be reliably obtained.
% The parameters of the effective Lagrangian are subject of the renormalization group equations. and below the heavy neutrino mass threshold they are traded for the Weinberg operators.
%As a result of our work
In what follows, we will show that, although the model has a rather limited parameter space, there is a phenomenologically acceptable parameter setting.

\section{Low-scale seesaw mechanisms and motivation of the Model}
%\section{Low-scale seesaw mechanisms and general concept of the Model}
%\section{Low-scale seesaw mechanisms and general goals of the Model}
%\section{Low-scale seesaw mechanisms and general idea of the Model}
\label{sec:Low-scaleSeesaw}
%
%%The neutrino masses, $m_\nu$, are at least six orders of magnitude smaller than the mass of the lightest charged fermion, and twelve orders of magnitude smaller than the electroweak scale $v$. Na\"\i vely and in correspondence with charged fermions of the Standard Model (SM), one would expect neutrino masses to be Dirac masses $m_D$ of not so an extremely different order of magnitude from the electroweak scale $v$. Nature however seems to choose an enormous hierarchy instead. If one holds onto the correspondence with the charged fermions, unnaturally small Yukawa parameters have to be introduced by hand. That can be avoided by means of the seesaw mechanism, which uses rather natural values for the involved quantities. It consists of the introduction of a seesaw scale, $M_R$, which is not protected by any of the SM symmetries and thus it may be arbitrarily larger than the electroweak scale. We can thus choose $M_R\gg m_D\sim v$. As a result, on the one hand $M_R$ determines the masses of heavy neutrino states, and on the other hand it provides a suppression factor $m_D/M_R$ for the light neutrino masses.
%
%%
In this section we want to introduce the low-scale seesaw mechanism and motivate our model, which will be presented in detail in the next section.

The conventional seesaw mechanism of type~I contains a neutrino Dirac mass $m_D$ coming from a Yukawa interaction with the Higgs field along with the electroweak symmetry breaking. In that case the seesaw scale is given by a right-handed Majorana neutrino mass $M_R$. Then the smallness of the neutrino mass relies on the suppression factor $m_D/M_R$. In contrast, the low-scale seesaw mechanism operates also with a small mass scale $\mu$, allowing for an additional suppression factor $\mu/m_D$, which relaxes the requirement on the seesaw scale $M_R$ to be extremely large, so that it may be not far above the
reach of current high energy experiments.
%lower limits.

The low-scale seesaw mechanism used in this work, limited here to a single generation, is built by introducing two right-handed neutrino fields, $\nu_{R}, S_{R}$, which are sterile under the SM gauge group and by assuming a neutrino mass matrix of the form:
\begin{equation}\label{M}
M_\nu = \beginm{ccc} 0 & m_D & \mu_\mathrm{lin} \\ m_D & \mu'_\mathrm{inv} & M_R \\ \mu_\mathrm{lin} & M_R & \mu_\mathrm{inv} \endm \,
\end{equation}
written in the basis $(\nu_L, \ \nu^{c}_R, \ S^{c}_R)$.
%Its matrix elements are $3\times 3$ matrices in the lepton flavor space, and
The $\nu_{L}-\nu_{L}^\C$ element vanishes by the electroweak gauge symmetry. We shall
set $\mu'_\mathrm{inv}=0$ because, as it is argued in Appendix~\ref{AppM}, for our purposes it is not a phenomenologically significant parameter. By setting either $\mu_\mathrm{inv/lin}=0$ the linear/inverse seesaw scenarios are obtained, respectively.

To obtain one light $\nu$ and two heavy $N_{\pm}$ seesaw neutrino mass eigenstates, the following hierarchy is usually assumed:
\begin{equation}\label{seesaw_hierarchyx}
\mu_\mathrm{lin},\mu_\mathrm{inv}\ll m_D\ll M_R \,.
\end{equation}
One of the conclusions of our analysis is that we should end up with a slightly different hierarchy, namely:
\begin{equation}\label{seesaw_hierarchy}
\mu_\mathrm{lin}\ll\mu_\mathrm{inv}\sim m_D\ll M_R \,,
\end{equation}
which however still provides a low-scale seesaw mechanism.
By diagonalization of the neutrino mass matrix \eqref{M} with $\mu'_\mathrm{inv}=0$, the light and heavy neutrino masses are obtained\footnote{More detailed expressions for mass eigenvalues of $M_\nu$ are given in Eq.~\eqref{mnu_App}.}
\begin{eqnarray}
m_{\nu} &\simeq& \mu_{\rm inv} \frac{m_D^2}{M_{R}^2}-2\mu_{\rm lin} \frac{m_D}{M_R} \,,\label{mnu}\\
m_{N_\pm} &\simeq&  M_R\pm\frac{1}{2}\mu_\mathrm{inv}\label{mN}
\end{eqnarray}

The lepton number assignment for the right-handed neutrino fields has a one-parameter freedom. There is a special assignment:
\begin{equation}
	L(\nu_{r}) = -L(S_{R}) = 1\,,
\end{equation}
in which the mass $M_R$ is lepton number invariant. The only LNV mass parameters in \eqref{M} are $\mu_\mathrm{lin}$ and $\mu_\mathrm{inv}$ ($\mu'_\mathrm{inv}$). If they are introduced via soft terms in the Lagrangian, then their values are protected by lepton number symmetry from acquiring large radiative corrections, i.e., their smallness is technically natural, and the necessary seesaw hierarchy \eqref{seesaw_hierarchyx} is preserved. Within our model, the LNV mass parameters appear dynamically, as they will be proportional to VEVs of composite scalar fields.
%As such they are not protected by the lepton number symmetry.
Their smallness must result from the details of the underlying dynamics, which are, at this stage, not fully specified but just parametrized as four-fermion interactions.

\section{Model setup}
\label{sec:ModelSetup}

We propose an extension of the SM with two sterile fermions $\nu_{R}$ and $S_{R}$, dubbed right-handed neutrinos, which form - with each other and with the SM leptons - composite scalar bosons via four-fermion interactions.
In the present paper we limit ourselves to only one generation of fermions in order to develop the formalism and test the key phenomenological features of the model. The study of flavor physics in this framework is considered the next step, to be made elsewhere.

%{\color{red}
%Even though we have decided not to consider the $\mu'_\mathrm{inv}$ mass parameter, in this section we introduce the model in a general way that keeps track of it. In the following sections we will not consider it any more.  DO WE NEED TO SAY THIS HERE? IT WILL FOLLOW NATURALLY. MAYBE WE ARE OVERLOADING THE TEXT WITH TOO MANY CROSSED IDEAS.
%}

\subsection{Effective theory description in terms of elementary fields}
\label{sec:EffTHElementary}
%Our model is then defined by the following Lagrangian:
At the level of elementary fields, the Lagrangian in our model is given by
%\begin{eqnarray}
%\eL &=& \eL_\mathrm{SM} + \eL_H + \eL_\mathrm{0} + \eL_R + \eL_\mathrm{lin} + \eL_\mathrm{inv} .
%\end{eqnarray}
%Here $\eL_\mathrm{0}$ corresponds to the kinetic terms of $\nu_R$ and $S_R$. In order to show the part of the Lagrangian relevant for our analysis we write the Lagrangian once again as
\begin{eqnarray}\label{L_model}
\eL &=& \eL_\mathrm{SM}^\prime + D^\mu H^\dag D_\mu H - \V(H^\dag H) - (y_H\overline{\ell_L} \tilde{H} \nu_R +\hc) \\
&& + \im\bar{\nu}_R\sles{\partial}\nu_R + \im\bar{S}_R\sles{\partial}S_R - (\overline{S^{c}_R} M_R \nu_R +\hc) \nonumber\\
&& - G_\mathrm{lin}(\bar{\ell}_L S_R)(\overline{S_R} \ell_L) - G_\mathrm{inv}(\overline{S^{c}_R} S_R)(\overline{S_R} S^{c}_R)
- G'_\mathrm{inv}(\overline{\nu^{c}_R} \nu_R)(\overline{\nu_R} \nu^{c}_R) \,
.\nonumber
\end{eqnarray}
Here $\eL_\mathrm{SM}^\prime$ is the single-family SM Lagrangian, from which we have pulled out the gauge-kinetic term of the Higgs field and the standard Higgs potential $\V(H^\dag H)$, characterized by its parameters $\mu_H$ and $\lambda_H$.

%\section{Symmetries of the model}

The Lagrangian \eqref{L_model} is SM gauge-invariant and has a global lepton number symmetry $U(1)_{L}$,
%\begin{equation}
%\mathcal{G}_\mathrm{local}=\SU{3}_c\times\SU{2}_L\times\U{1}_Y
%\end{equation}
%and conserves the global lepton and baryon numbers
%\begin{equation}
%\mathcal{G}_\mathrm{global}=\U{1}_B\times\U{1}_L\,.
%\end{equation}
with the field assignment shown in
%The quantum number assignment to the sector relevant for neutrino mass generation is shown in
Table \ref{Qno}.

\begin{table}[h]
\begin{center}
\begin{tabular}{c|cccc|ccc}
%$\mathcal{G}$
Group
& $\ell_L$ & $H$ & $\nu_R$ & $S_R$ & $\Sigma$ & $\Phi$ & $\Phi'$  \\
\hline
\hline
$\U{1}_Y$ & $-1$ & $+1$ & $0$ & $0$ & $-1$ & $0$ & $0$ \\
$\U{1}_L$ & $+1$ & $0$ & $+1$ & $-1$ & $+2$ & $-2$ & $+2$ \\
\hline
$\SU{2}_L$ & $\mathbf{2}$ & $\mathbf{2}$ & $\mathbf{1}$ & $\mathbf{1}$ & $\mathbf{2}$ & $\mathbf{1}$ & $\mathbf{1}$ \\
\hline
\hline
\end{tabular}
\end{center}
\caption[]{\small The SM gauge group and lepton number assignments for
%Quantum number assignment to
the fields relevant for neutrino mass generation. }
\label{Qno}
\end{table}

The global symmetries encounter the same axial anomalies as in  the SM, generated by non-perturbative effects of the electroweak and QCD gauge dynamics. Therefore  $B-L$ remains an exact symmetry.
%: -1 -- as an exact symmetry.

%\section{Condensation scheme and energy scales}

We consider our model, defined by the Lagrangian $\eL$ in Eq.~\eqref{L_model}, as an effective %low-scale
description of some more fundamental underlying theory at higher energy scales.
%This underlying physics presumably exhibits some symmetry breaking and develops a spectrum of mass parameters.
Moreover, we assume that the field content of such an underlying theory can be divided into a heavy and a light sector, and for the characteristic scale of the heavy sector, $M_\mathrm{heavy}$,  we assume
%contains fields whose masses have a
%characteristic magnitude denoted by $M_\mathrm{heavy}$,  with
$M_\mathrm{heavy}<\Lambda_\mathrm{Planck}$, in order to not
%interfere with
reach
quantum gravity effects. The light sector consists of the fields participating in the Lagrangian \eqref{L_model}, which are all massless except for the elementary Higgs field with a $\mu_H$-mass parameter in the Higgs  potential and the right-handed neutrino fields $\nu_{R}$, $S_{R}$ with their mass parameter $M_R$.
%
%We assume that integrating out
The heavy sector of the underlying theory is integrated out and assumed to generate  the four-neutrino interactions in Eq.~\eqref{L_model},
%$\eL_\mathrm{lin}$ and $\eL_\mathrm{inv}$ are generated,
with the coupling constants of the order of
%given by
\begin{equation}
G_\mathrm{lin},G_\mathrm{inv}, G_\mathrm{inv}'\propto M_\mathrm{heavy}^{-2}\,.
\end{equation}
Thus,
%$G_{K}^{-1/2}$
\Sergey{$M_\mathrm{heavy}$}
is a cut-off scale of the effective theory defined by the Lagrangian \eqref{L_model}.

\subsection{Condensation and energy scales}
\label{sec:CondensationScheme}
The key point of our model, based on Lagrangian \eqref{L_model}, is the assumption that due to the attractiveness of the four-fermion interactions, the following
\Sergey{scalar}
bound states of fermion pairs are formed:
\begin{subequations}\label{bound_states}
	\begin{eqnarray}
	\Sigma &\sim & (\overline{S_R} \ell_L) \,,\\
	\Phi &\sim & (\overline{S^{c}_R} S_R) \,,\\
	\Phi' &\sim & (\overline{\nu^{c}_R} \nu_R) \,,
	\end{eqnarray}
\end{subequations}
at some scale $\Lambda < M_\mathrm{heavy}$.
\Sergey{At this same scale $\Lambda$ or somewhere below}
%: -2 -- Below the scale $\Lambda$
the composite bound states develop non-trivial VEVs, corresponding to the fermion-pair condensates
% in the corresponding channel:
\begin{subequations}\label{vevs}
	\begin{eqnarray}
	\langle\Sigma\rangle &\equiv& \frac{1}{\sqrt{2}}\beginm{c}v_\Sigma \\ 0\endm \sim  \langle\overline{S_R} \ell_L\rangle \,,\\
	\langle\Phi\rangle &\equiv& \  \  \frac{v_\Phi}{\sqrt{2}}  \quad \sim \  \langle\overline{S^{c}_R} S_R\rangle \,,\\
	\langle\Phi'\rangle &\equiv& \  \  \frac{v_{\Phi'}}{\sqrt{2}}  \quad \sim \  \langle\overline{\nu^{c}_R} \nu_R\rangle \,.
	\end{eqnarray}
\end{subequations}
%:{SK} How this is related with the gap equation?
%{\color{blue}The condensation conditions will be addressed in next sections.}
%
In what follows we shall call $\Lambda$ \textit{the compositeness scale}.
%{\color{red}  Does it coincide with the bound-state formation??? See below (\ref{bound_states}).}
The VEVs $v_\Sigma$, $v_\Phi$ and $v_\Phi'$ break $U(1)_{L}$, generating
\Sergey{the $\Delta L =2$}
%: -3-- the LNV
entries of the mass matrix \eqref{M} of the neutrino sector, so that
\begin{eqnarray}
%	m_D & = & \frac{y_H v_H}{\sqrt{2}}\,,\label{mD}\\
\mu_\mathrm{lin} & = & \frac{y_\Sigma v_\Sigma}{\sqrt{2}}\,,\label{mulin}\\
\mu_\mathrm{inv} & = & \frac{y_\Phi v_\Phi}{\sqrt{2}}\, ,\label{muinv}\\
\mu'_\mathrm{inv} & = & \frac{y_{\Phi'} v_{\Phi'}}{\sqrt{2}}\,.\label{muinv-1}
\end{eqnarray}
The effective Yukawa couplings $y_{\Sigma}, y_{\Phi}$ and $y_{\Phi'}$ stem from the four-fermion couplings
$G_{\rm lin}, G_{\rm inv}$ and $G'_{\rm inv}$ in Eq.~\eqref{L_model}, and the corresponding relations between these Yukawas and four-fermion couplings will be discussed in the next sections.
%:....As we already noted in

The Dirac type entry is as usual
\begin{equation}
	m_D = \frac{y_H v_H}{\sqrt{2}}\,,\label{mD}
\end{equation}
where $v_H$ is the VEV
%$v_H$ is
developed by the elementary electroweak doublet Higgs field
%: -4-- where the vacuum expectation value $v_H$ is developed by the elementary Higgs field $H$
\begin{eqnarray}\label{vevH}
\langle H\rangle = \frac{1}{\sqrt{2}}\beginm{c}0\\v_H\endm\,,
\end{eqnarray}
according to its potential, which at tree level is not affected by the right-handed neutrino condensation and is governed mainly by the usual SM parameters $\mu_H$ and $\lambda_H$.

For our model it is crucial that the right-handed neutrinos are not too heavy to be integrated out before their condensation happens.
%An important field-theoretical restriction of our model is the so called non-decoupling condition. %Basically it reflects our requirement that the right-handed neutrinos are not too heavy to be %integrated out before their condensation happens.
%
Therefore we require the non-decoupling condition:
% in the form:
\begin{equation}\label{nondecoupling}
M_R < \Lambda \,.
\end{equation}

%Considering the parameter
For simplicity we will consider $\mu'_\mathrm{inv}=0$ in the rest of this work, which can be interpreted as the fact that the corresponding four-fermion coupling constant $G'_{\rm inv}$ is sub-critical so that $v_\Phi'=0$. Moreover, we will take the even stronger assumption that $G'_{\rm inv}$ is so weak that not even the bound state $\Phi'$ is formed.

\section{Effective description of the % right-handed
neutrino condensation}
\label{sec:EffectiveDescriptionNuCond}

Here we  use
% develop
the formalism developed in Refs.~\cite{Bardeen:1989ds,Hill:1990ge}, which is suitable for the realization of the above-described scenario of neutrino condensation, starting from the attractive four-fermion interactions in the
%``fundamental''
Lagrangian \eqref{L_model} of our model. Akin to the famous case of condensed matter physics, where the electron-pair condensation is preceded by Cooper-pairing, in our model the fermion-antifermion condensation inevitably involves formation of bound states, which in an effective theory are described by the corresponding effective bosonic fields, which in turn introduce new phenomenology at low energies. We start with the definition of this effective bosonized theory.

\subsection{Bosonization}

As we already discussed, the key hypothesis of our model is that the attractive four-fermion interactions in Eq.~\eqref{L_model} lead to the formation of bound states at the compositeness scale $\Lambda$, according to Eq.~\eqref{bound_states}. This non-perturbative phenomenon can be suitably described by the bosonization prescription\footnote{This is a particular case of the Hubbard--Stratonovich transformation. For a discussion see for, instance,
%: {SK 25.12.2018-1} More references are needed here.
Ref.~\cite{Cvetic:1997eb}.}
introducing the auxiliary fields $\Sigma_\Lambda$ and $\Phi_\Lambda$:
\begin{eqnarray}
\label{eq:Auxiliary-Habbard-Stat-1}
\eL_{\mathrm{lin};\Lambda} &=& - (y_{\Sigma;\Lambda}\bar{\ell}_L\Sigma_\Lambda S_R + \hc) - \mu^{2}_{\Sigma;\Lambda}\Sigma_{\Lambda}^\dag\Sigma_\Lambda \,, \\
\label{eq:Auxiliary-Habbard-Stat-2}
\eL_{\mathrm{inv};\Lambda} &=& - (y_{\Phi;\Lambda}\overline{S_R}\Phi_\Lambda S^{c}_R + \hc) - \mu^{2}_{\Phi;\Lambda}\Phi_{\Lambda}^\dag\Phi_\Lambda \,.
\end{eqnarray}
The auxiliary fields  have no kinetic terms and as such can be eliminated from the Lagrangian by means of their non-dynamical equations of motion
\begin{eqnarray}
\label{eq:Sigma-1}
\Sigma_{\Lambda}^\dag &=& \frac{y_{\Sigma;\Lambda}}{\mu^{2}_{\Sigma;\Lambda}}\bar{\ell}_L S_R \,, \\
\label{eq:Phi-1}
\Phi_{\Lambda}^\dag &=& \frac{y_{\Phi;\Lambda}}{\mu^{2}_{\Phi;\Lambda}}\overline{S_R}S^{c}_R  \, ,
\end{eqnarray}
and their respective hermitian conjugates \Adam{for} $\Sigma_{\Lambda}$ and  $\Phi_{\Lambda}$. We thus recover the original four-fermion interactions from Eq.~\eqref{L_model} by identifying
the coefficients:
\begin{subequations}\label{G_lin_inv}
\begin{eqnarray}
G_{\mathrm{lin}} &=& \frac{y_{\Sigma;\Lambda}^2}{\mu^{2}_{\Sigma;\Lambda}} \,, \\
G_{\mathrm{inv}} &=& \frac{y_{\Phi;\Lambda}^2}{\mu^{2}_{\Phi;\Lambda}} \,.
\end{eqnarray}
\end{subequations}

\subsection{Effective low-energy Lagrangian}

The bosonized model Lagrangians \eqref{eq:Auxiliary-Habbard-Stat-1} and \eqref{eq:Auxiliary-Habbard-Stat-2} evolve from the scale $\Lambda$ down to some low-energy scale $m$,
\Sergey{in accordance with
%according to
the corresponding Renormalization Group Equations (RGEs),}
%according to the equations of Renormalization Group Equation (RGE),
\begin{equation}
\label{eq:Lagrangian-RGE}
\eL_\Lambda\rightarrow \eL_m \,.
\end{equation}
\Sergey{As it is well known, the RGE} evolution can be interpreted as integrating out the higher-energy field modes in the interval $(m,\Lambda)$. This
\Adam{procedure}
%evolution
%results in
leads to
%the creation
%the generation
the appearance
of effective operators generated by quantum corrections, such as the kinetic terms of the composite fields $\Phi$ and $\Sigma$, which are weighed by the wave function renormalization coefficients $Z_{\Sigma,\Phi;m}$, whose main radiative one-loop contribution comes from the Yukawa interaction
\begin{equation}\label{Z}
	Z_{\Sigma,\Phi;m} = \frac{y_{\Sigma,\Phi;\Lambda}^2}{(4\pi)^2}\ln\frac{\Lambda}{m}\ \sim\ \mathcal{O}(1) \ \stackrel{m\rightarrow\Lambda}{\longrightarrow}\ 0\,.
\end{equation}
The coefficients $Z_{\Sigma,\Phi;m}$ vanish in the limit $m\rightarrow\Lambda$, since the kinetic operators are not present in the Lagrangian \eqref{eq:Auxiliary-Habbard-Stat-1} and \eqref{eq:Auxiliary-Habbard-Stat-2}, relevant for the scale $\Lambda$. On the other hand, the mass term operators are present at the scale $\Lambda$, and for $m<\Lambda$ their coefficients $\mu^{2}_{\Sigma,\Phi;m}$ only get radiative corrections from the Yukawa interaction according to
\begin{equation}
\mu^{2}_{\Sigma,\Phi;m} = \mu^{2}_{\Sigma,\Phi;\Lambda}-\frac{2y_{\Sigma,\Phi;\Lambda}^2}{(4\pi)^2}(\Lambda^2-m^2) \ \stackrel{m\rightarrow\Lambda}{\longrightarrow}\ \mu^{2}_{\Sigma,\Phi;\Lambda}\,. \label{mu2}
\end{equation}
Actually,
\Sergey{it is not necessary to perform the above-mentioned integrating-out of the high-energy modes}
%the integration
%does not need to be performed
explicitly: it is sufficient to realize that all the operators allowed by the symmetries
of the initial Lagrangian
(\ref{eq:Auxiliary-Habbard-Stat-1}), (\ref{eq:Auxiliary-Habbard-Stat-2})
will be
%RGEs-generated
radiatively generated
and contribute to the effective Lagrangian.
%Taking into account just
Taking into account only
the relevant operators, we end up after the necessary field normalization, with a renormalizable effective theory valid at scales lower than $\Lambda$, described by the effective Lagrangian
\begin{eqnarray}\label{L}
\eL_\mathrm{eff} &=& \eL_\mathrm{SM}^\prime + D^\mu H^\dag D_\mu H + D^\mu \Sigma^\dag D_\mu \Sigma + \partial^\mu \Phi^\dag \partial_\mu \Phi - \V_\mathrm{eff}(H,\Sigma,\Phi) \nonumber\\
&& - (y_H\overline{\ell_L} \tilde{H} \nu_R + y_\Sigma\bar{\ell}_L\Sigma S_R + y_\Phi\overline{S_R}\Phi S^{c}_R +\hc) \\
&& + \im\bar{\nu}_R\sles{\partial}\nu_R + \im\bar{S}_R\sles{\partial}S_R - (\overline{S^{c}_R} M_R \nu_R +\hc) \,,\nonumber
\end{eqnarray}
where
\begin{eqnarray}\label{V}
\V_\mathrm{eff}(H,\Sigma,\Phi) &=& \mu_{H}^2H^\dag H + \mu_{\Sigma}^2\Sigma^\dag\Sigma + \mu_{\Phi}^2\Phi^\dag\Phi \\
&& + \frac{1}{2}\lambda_H(H^\dag H)^2 + \frac{1}{2}\lambda_{\Sigma}(\Sigma^\dag\Sigma)^2 + \frac{1}{2}\lambda_{\Phi}(\Phi^\dag\Phi)^2  \nonumber \\
&& + \lambda_{\Phi H}(\Phi^\dag\Phi)(H^\dag H) + \lambda_{\Phi\Sigma}(\Phi^\dag\Phi)(\Sigma^\dag\Sigma)  \nonumber \\
&& + \lambda_{H\Sigma}(H^\dag H)(\Sigma^\dag\Sigma) + \lambda'_{H\Sigma}(\Sigma^\dag\tilde{H})(\tilde{H}^\dag\Sigma)   \nonumber \\
&& + \Big[\kappa\,\Phi^\dag(H^\dag\tilde{\Sigma}) + \hc\Big] ,  \nonumber
\end{eqnarray}
is the effective potential for the scalar fields.

The parameters of the effective Lagrangian run according to their RGEs. We use one-loop RGEs %equations
%:- 8-- which can be found
\Sergey{given}
 in Appendix~\ref{AppRGE} \Adam{and proceed
% along
% on
\Adam{with}
 the standard strategy given, e.g., in \cite{Hill:1990ge,Cvetic:1997eb,Antusch:2002xh}}.
%
 %:{SK 25.12.2018-2}
% The origin of the dynamically generated parameters $y_{\Sigma}$, $y_{\Phi}$, $\lambda_{\Sigma}$, $\lambda_{\Phi}$, $\lambda_{H\Sigma}$, $\lambda'_{H\Sigma}$, $\lambda_{\Phi H}$, $\lambda_{\Phi\Sigma}$ and $\kappa$ is reflected in their RGE-running by specific boundary conditions at the scale $\Lambda$.
\Sergey{The parameters $y_{\Sigma}$, $y_{\Phi}$, $\lambda_{\Sigma}$, $\lambda_{\Phi}$, $\lambda_{H\Sigma}$, $\lambda'_{H\Sigma}$, $\lambda_{\Phi H}$, $\lambda_{\Phi\Sigma}$ and $\kappa$ are dynamically generated by their RGE running from specific boundary conditions\Adam{, which follow from the matching of the Lagrangians $\eL_m$ [Eq.~\eqref{eq:Lagrangian-RGE}] and $\eL_\mathrm{eff}$ [Eq.~\eqref{L}] at the compositeness} scale $\Lambda$, }
and from the renormalization of the $\eL_m$ parameters $Z_{\Sigma;m}$, $Z_{\Phi;m}$, $y_{\Sigma;m}$, $y_{\Phi;m}$, $\lambda_{\Sigma;m}$, $\lambda_{\Phi;m}$, $\lambda_{H\Sigma;m}$, $\lambda'_{H\Sigma;m}$, $\lambda_{\Phi H;m}$, $\lambda_{\Phi\Sigma;m}$ and $\kappa_m$ shown, e.g., in Eq.~\eqref{Z}. From the relation between the Lagrangians $\eL_m$ and $\eL_\mathrm{eff}$, which is given by the proper normalization of the kinetic terms of the composite scalar fields
\begin{eqnarray}\label{eq:renormalization-1}
\eL_\mathrm{eff}(\Sigma,\Phi)=\eL_m(\Sigma_m\rightarrow\Sigma/\sqrt{Z_{\Sigma;m}}\,,\,\Phi_m\rightarrow\Phi/\sqrt{Z_{\Phi;m}}),
\end{eqnarray}
we obtain the relations
\begin{eqnarray}
\label{eq:renormalization-param-1}
&\mu_{\Sigma}^2=\frac{\mu_{\Sigma;m}^2}{Z_{\Sigma;m}}\,,\,\mu_{\Phi}^2=\frac{\mu_{\Phi;m}^2}{Z_{\Phi;m}}\,,& \\
&y_{\Sigma}=\frac{y_{\Sigma;m}}{\sqrt{Z_{\Sigma;m}}}\,,\,y_{\Phi}=\frac{y_{\Phi;m}}{\sqrt{Z_{\Phi;m}}}\,,\kappa=\frac{\kappa_m}{\sqrt{Z_{\Sigma;m}Z_{\Phi;m}}}\,,& \\
&\lambda_{\Sigma}=\frac{\lambda_{\Sigma;m}}{Z_{\Sigma;m}^2}\,,\,\lambda_{\Phi}=\frac{\lambda_{\Phi;m}}{Z_{\Phi;m}^2}\,,\,\lambda_{H\Sigma}=\frac{\lambda_{H\Sigma;m}}{Z_{\Sigma;m}}\,,\,&\\
&\lambda'_{H\Sigma}=\frac{\lambda'_{H\Sigma;m}}{Z_{\Sigma;m}}\,,\,\lambda_{\Phi H}=\frac{\lambda_{\Phi H;m}}{Z_{\Phi;m}}\,,\,\lambda_{\Phi\Sigma}=\frac{\lambda_{\Phi\Sigma;m}}{Z_{\Sigma;m}Z_{\Phi;m}}\,.&
\end{eqnarray}
%: -9--and from there
\Sergey{
From these relations we get the boundary conditions at the matching scale $\Lambda$}
\begin{eqnarray}
	y_{\Sigma}\,,\,y_{\Phi}&\stackrel{m\rightarrow\Lambda}{\longrightarrow}&\infty\,,\label{Yukawa_cond}\\
	\frac{\lambda_{\Sigma}}{y_{\Sigma}^4}\,,\,\frac{\lambda_{\Phi}}{y_{\Phi}^4}\,,\,\frac{\lambda_{H\Sigma}}{y_{\Sigma}^2}\,,\,\frac{\lambda'_{H\Sigma}}{y_{\Sigma}^2}\,,\,\frac{\lambda_{\Phi H}}{y_{\Phi}^2}\,,\,\frac{\lambda_{\Phi\Sigma}}{y_{\Sigma}^2y_{\Phi}^2} &\stackrel{m\rightarrow\Lambda}{\longrightarrow}&0\,,
\label{lambda_cond}\\
	\frac{\kappa}{y_{\Sigma}y_{\Phi}}&\stackrel{m\rightarrow\Lambda}{\longrightarrow}&0\,.\label{kappa_cond}
\end{eqnarray}

Notice that the boundary conditions for the Yukawa parameters \eqref{Yukawa_cond} exhibit an ill behavior. When approaching
%condensation
%{\color{blue} compositeness}
the matching scale $\Lambda$ from below, these Yukawa parameters grow, indicating
%This signals their dynamical origin.
\Sergey{ their non-perturbative origin. In fact, these couplings appear as a result of the formation of the bound states $\Sigma$ and $\Phi$ at the scale $\Lambda$, which is essentially a non-perturbative phenomenon.
%Above some value of the Yukawa parameters  the effective theory is entering a highly non-perturbative regime and
%Note that in the approach we are working in the matching scale $\Lambda$ coincides with the scale of the bound state (\ref{eq:Sigma-1}), (\ref{eq:Phi-1}) formation. To wit, $\Lambda$ is a compositeness scale.
 Therefore, once $y_{\Sigma}\,,\,y_{\Phi}$ become larger than some value -- typically $4\pi$ -- the perturbative one-loop RGEs cannot be trusted anymore. In practice this means that we have lost the relation between $y_{\Sigma}\,,\,y_{\Phi}$ of the effective theory (\ref{L}) and the four-fermion couplings in Eqs.~(\ref{G_lin_inv}) of the underlying theory (\ref{L_model}).} Consequently, instead of using the ill defined matching condition we follow the standard strategy described in e.g. \cite{Hill:1990ge,Cvetic:1997eb,Antusch:2002xh}, and for
 %practical purposes
 ease of numerical calculations we set
\begin{eqnarray}
\label{Yuk_bound}
&&y_\Sigma(\Lambda)=y_\Phi(\Lambda)= y_{0}\,,
\end{eqnarray}
where $y_{0}$ is some finite value, typically $\sim4\pi$.
%
%Therefore we set $y_\Sigma$ and $y_\Phi$ at the scale $\Lambda$ conventionally and deliberately to finite values
%%\begin{subequations}\label{Yuk_bound}
%\begin{eqnarray}
%\label{Yuk_bound}
%y_\Sigma(\Lambda)=3\,,\,&&y_\Phi(\Lambda)= 3\,.
%\end{eqnarray}
%%\end{subequations}
%The point is that, anyway, the actual low-energy values are only very weakly sensitive to the high-energy values. Typically, they run from their initial value at the high energy scale down to values $\sim1$ at scales $\sim M_R$.
%
As will be shown below, such arbitrariness in the boundary condition is justified by the fact that the low-energy values  are only very weakly sensitive to the high-energy values of $y_{\Sigma}\,,\,y_{\Phi}$.
%Typically, they run from their initial value at the high energy scale down to values $\sim1$ at scales $\sim M_R$.
%$y_\Sigma$ and $y_\Phi$ at the scale $\Lambda$ conventionally and deliberately to finite values
%\begin{subequations}\label{Yuk_bound}

%\end{subequations}
%

From  \eqref{lambda_cond}, \eqref{kappa_cond} and \eqref{Yuk_bound} we obtain the boundary conditions for the rest of the parameters:
\begin{eqnarray}
\lambda_{K}(\Lambda)&=&0\,,\ \ K=\Sigma,\Phi,H\Sigma,H\Sigma',\Phi H,\Phi\Sigma\,,\label{lambda_bound}\\
\kappa(\Lambda)&=&0\,.\label{kappa_bound}
\end{eqnarray}

The effective scalar potential $\V_\mathrm{eff}(H,\Sigma,\Phi)$ in Eq.~\eqref{V} has a non-trivial minimum, which defines the symmetry breaking pattern. As in the SM, we assume $\mu_{H}^2<0$. In order that both LNV
%vacuum expectation values,
VEVs,
$v_\Sigma$ and $v_\Phi$, have nonzero values, $\mu_{\Sigma}^2$ and $\mu_{\Phi}^2$ must be negative.

Now, in order to set up a low-scale
%sake of
seesaw mechanism, both $\mu_{\mathrm{lin}}$ and $\mu_{\mathrm{inv}}$ have to be small compared to the other neutrino mass parameters. Since the Yukawa coupling parameters in \eqref{mulin} and \eqref{muinv} do not help guaranteeing  this smallness, because $y_{\Sigma,\Phi}\sim1$, the VEVs $v_\Sigma$ and $v_\Phi$ must be small, which requires the smallness of $|\mu_{\Sigma}^2|$ and $|\mu_{\Phi}^2|$ at low scales $m\rightarrow0$.
Accordingly, Eqs.~\eqref{G_lin_inv} and \eqref{mu2}  provide a requirement on the underlying four-fermion interaction parameters $G_\mathrm{lin}$ and $G_\mathrm{inv}$, which have to be tuned to be just slightly super-critical:
\begin{equation}
	\mathrm{to\ obtain}\ \ |\mu_{\Sigma,\Phi}^2|\ll\Lambda^2\ :\ \ \ 0<\frac{G_{\mathrm{lin},\mathrm{inv}}}{G_{\mathrm{lin},\mathrm{inv}}^\mathrm{crit}}-1\ll 1
\end{equation}
where $G_{\mathrm{lin},\mathrm{inv}}^\mathrm{crit}\equiv8\pi^2/\Lambda^2$ is the critical value of the four-fermion coupling parameters. We do not try to explain this feature of the underlying new dynamics in this work, but keep it as one of the subjects for future work.

\Sergey{Interestingly, our model allows for a unique  triple scalar coupling in Eq.~(\ref{V}) with the calculable constant $\kappa$. As will be shown in what follows, this triple coupling plays an essential role in order to meet all the phenomenological constraints.
%, as we will show in the rest of the work.
}Although the coupling constant $\kappa$ is in general complex, its phase can be absorbed into the redefinition of the field $\Phi$. Therefore, all the coupling constants in the effective potential are real parameters. This phase absorption corresponds to the known fact that in the scalar sector of a model with two Higgs doublets and one complex Higgs singlet, there is no source of CP violation \cite{IVANOV2017160}.

\subsection{Generalized Weinberg operators}

The right-handed neutrino mass $M_R$ is the highest mass scale in our model, and below this scale the right-handed neutrinos decouple from \Sergey{the low-energy observables.}
%the model.
As a consequence, below $M_R$ all three neutrino Yukawa interactions weighed by the Yukawa coupling parameters $y_H$, $y_\Sigma$ and $y_\Phi$, are traded for the effective operators of higher dimensions that result from integrating out the right-handed neutrinos. The part of the Lagrangian containing these effective operators is
\begin{equation}\label{Lw}
\eL_w = \frac{w_\mathrm{inv}}{2M_{R}^2}(\bar{\ell}_L\tilde H)\Phi(H^\dag\ell_{L}^{c}) + \frac{w_\mathrm{lin}}{M_{R}}(\bar{\ell}_L\Sigma)(H^\dag\ell_{L}^{c}) +\hc
\end{equation}
We will refer to these operators as \textit{generalized Weinberg operators}, where $w_\mathrm{inv}$ and $w_\mathrm{lin}$ are dimensionless Weinberg parameters.
After the scalar fields develop their VEVs, these terms will directly provide the Majorana mass term for the active neutrino
\begin{equation}
\eL_{m_\nu} = \frac{1}{2}m_\nu(\bar{\nu}_L\nu_{L}^{c})+\hc\,,
\end{equation}
where $m_\nu$ is obtained from $\eL_w$ in Eq.~\eqref{Lw} as:
\begin{equation}\label{mnuw}
m_\nu = w_\mathrm{inv}\frac{v_{H}^2v_\Phi}{2\sqrt{2}M_{R}^2} + w_\mathrm{lin}\frac{v_H v_\Sigma}{2M_{R}} \,.
\end{equation}
On the other hand, calculating the same $m_\nu$ from $\eL_\mathrm{eff}$ in Eq.~\eqref{L},  we obtain an expression for the neutrino mass in terms of the Yukawa couplings:
\begin{equation}
m_\nu = y_{H}^2y_\Phi\frac{v_{H}^2v_\Phi}{2\sqrt{2}M_{R}^2} - y_H y_\Sigma\frac{v_H v_\Sigma}{M_{R}} \,.
\end{equation}
This leads to the matching condition at the scale $M_R$
\begin{eqnarray}
\left. w_\mathrm{inv}\right|_{m=M_R} &=& \left. y_{H}^2y_\Phi\right|_{m=M_R} \,, \label{wyHyF}\\
\left. w_\mathrm{lin}\right|_{m=M_R} &=& \left. -2y_{H}y_\Sigma\right|_{m=M_R} \,. \label{wyHyS}
\end{eqnarray}
Introducing the generalized Weinberg operators has the advantage of allowing us to avoid the procedure of diagonalizing the neutrino mass matrix \eqref{M} in order to determine the light neutrino mass, which
would require inserting
%feed
the low-energy values of the Yukawa coupling parameters $y_{H,\Sigma,\Phi}(m_\mathrm{low})$ into the entries of the neutrino mass matrix \eqref{M}. In principle, the low-energy scale, at which the neutrino mass is determined, $m_\mathrm{low}$, should be taken of the same order of magnitude as the neutrino mass itself $m_\mathrm{low}\sim m_\nu$. This would, however, entale a trouble, because the one-loop RGEs \Sergey{
drive
%guide
the Yukawa couplings to
large
%higher beyond-perturbative and unphysical
non-perturbative values at such small energy scale. This problem is
%removed
eliminated
by trading, at the $M_R$-scale, the Yukawa couplings for the Weinberg parameters, whose RGE running is safe all the way down to arbitrarily low scales.}

\subsection{Minimization of the effective potential and symmetry breaking}

The minimum of the effective potential $\V_\mathrm{eff}(H,\Sigma,\Phi)$ in Eq.~(\ref{V}) determines the values of the VEVs \eqref{vevs} and \eqref{vevH} as a solution of the equations
\begin{eqnarray}
\frac{\partial}{\partial v_H} \V_\mathrm{eff}(\langle H\rangle,\langle\Sigma\rangle,\langle\Phi\rangle) &=& 0 \,, \\
\frac{\partial}{\partial v_\Sigma}\V_\mathrm{eff}(\langle H\rangle,\langle\Sigma\rangle,\langle\Phi\rangle) &=& 0 \,, \\
\frac{\partial}{\partial v_\Phi} \V_\mathrm{eff}(\langle H\rangle,\langle\Sigma\rangle,\langle\Phi\rangle) &=& 0 \,.
\end{eqnarray}
%They satisfy the equations
\Sergey{Explicitly we have}
%\begin{subequations}
\begin{eqnarray}
-\mu_{H}^2 &=& \frac{1}{2v_H}\Big[-\sqrt{2}\kappa v_\Sigma v_\Phi + \lambda_H v_{H}^3 +  (\lambda_{H\Sigma}+\lambda'_{H\Sigma})v_H v_{\Sigma}^2 + \lambda_{\Phi H} v_H v_{\Phi}^2 \Big] \,, \nonumber\\
-\mu_{\Sigma}^2 &=& \frac{1}{2v_\Sigma}\Big[-\sqrt{2}\kappa v_H v_\Phi + \lambda_\Sigma v_{\Sigma}^3 +  (\lambda_{H\Sigma}+\lambda'_{H\Sigma})v_\Sigma v_{H}^2 + \lambda_{\Phi \Sigma} v_\Sigma v_{\Phi}^2 \Big] \,, \nonumber\\
-\mu_{\Phi}^2 &=& \frac{1}{2v_\Phi}\Big[-\sqrt{2}\kappa v_H v_\Sigma + \lambda_\Phi v_{\Phi}^3 + \lambda_{\Phi H}v_{H}^2v_\Phi + \lambda_{\Phi \Sigma}v_{\Sigma}^2v_\Phi \Big] \label{mu}
\end{eqnarray}
%\end{subequations}
in accordance with Ref.~\cite{PhysRevD.86.115024}. These equations can be used to trade the $\mu_{H,\Sigma,\Phi}$ parameters for the VEVs  $v_\Sigma$, $v_\Phi$ and $v_H$ in the potential $\V_\mathrm{eff}(H,\Sigma,\Phi)$ of Eq.~\eqref{V}. From Eqs.~\eqref{mu} we can see that unless $\kappa=0$, the solution with all three VEVs $v_H$, $v_\Sigma$ and $v_\Phi$ non-zero is the only one available.

The stability of the vacuum is guaranteed by the positive definiteness of the Hessian matrix
\begin{eqnarray}
\mathcal{H}_{ij}\equiv\frac{\partial^2} {\partial v_i\partial v_j} \V_\mathrm{eff}(\langle H\rangle,\langle\Sigma\rangle,\langle\Phi\rangle)&& \ \ \mathrm{where} \ \ \ i,j=H,\Sigma,\Phi \,,
\end{eqnarray}
which is calculated by using Eqs.~\eqref{mu} and is actually equivalent to the scalar boson mass matrix, written explicitly in Eq.~\eqref{M_hHs}.

\section{Properties of Higgs bosons}

In order to derive
%To derive
the properties of the physical Higgs scalar excitations, we
%need to
shift their fields by their VEVs:
\begin{subequations}\label{shift}
\begin{eqnarray}
H &=& \beginm{c}a_{H}^+\\(v_H+h_H+\im a_{H})/\sqrt{2}\endm \,,\\
\Sigma &=& \beginm{c}(v_\Sigma+h_\Sigma+\im a_{\Sigma})/\sqrt{2} \\ a_{\Sigma}^- \endm \,,\\
\Phi &=& (v_\Phi+h_\Phi+\im a_{\Phi})/\sqrt{2} \,,
\end{eqnarray}
\end{subequations}
%By performing
\Sergey{
%this shift, our effective potential now depends on fields appropriate for the true ground state:
by which the effective potential (\ref{V}) becomes a function of the fields corresponding to the true ground state:}
\begin{equation}
\V_\mathrm{eff}(H,\Sigma,\Phi) \longrightarrow \V_\mathrm{eff}(\vec\phi) \,,
\end{equation}
where
\begin{equation}
\vec{\phi}\equiv(h_H,a_H,a_{H}^-,a_{H}^+,h_\Sigma,a_\Sigma,a_{\Sigma}^-,a_{\Sigma}^+,h_\Phi,a_\Phi) \,.
\end{equation}
\Sergey{
%Now, let us rewrite the effective potential (\ref{V}) and all the other Lagrangian terms in
The scalar field mass eigenstates are the eigenstates of the $10\times10$ matrix:
%As the next step, the effective potential and all the other terms of the Lagrangian should be written in the basis of the mass eigenstates. The mass eigenstates corresponds to the eigenvalues of the $10\times10$ matrix:
}
\begin{equation}\label{M_Higgs}
\Big[M_{\mathrm{Higgs}}^2\Big]_{ij}=\frac{\partial^2}{\partial \phi_{i}\partial\phi_{j}} \V_\mathrm{eff}\,,
\end{equation}
where
%the fields
$\phi_i$ are the components of
%the vector
the field $\vec{\phi}$.
\Sergey{
%The diagonalization matrix determines the admixture of the fields $\phi_i$ in the mass eigenstates.
The matrix which diagonalizes the above mass-squared matrix, determines the admixture of the fields $\phi_i$ in the corresponding mass eigenstates.}
Clearly, since the $C$ and $P$ symmetries are conserved within the Higgs boson sector, the mass matrix $M_{\mathrm{Higgs}}^2$
%disentangles
splits
into blocks of charged bosons $(a_{H}^-,a_{H}^+,a_{\Sigma}^-,a_{\Sigma}^+)$, pseudo-scalar bosons $(a_H,a_\Sigma,a_\Phi)$, and scalar bosons $(h_H,h_\Sigma,h_\Phi)$. \Sergey{More details
%of which we summarize
are given in the Appendix~\ref{App_Higgs}. }

\subsection{Higgs bosons mass spectrum and mixing}

\Sergey{
Diagonalizing the mass matrices \eqref{M_charged}, \eqref{M_pseudo} and \eqref{M_hHs},
we obtain the Higgs boson mass eigenstates, their masses and mixing.
%The Higgs boson mass eigenstates, their masses and mixing are obtained by the diagonalization of the mass matrices \eqref{M_charged}, \eqref{M_pseudo} and \eqref{M_hHs}.
Let us summarize their main features.}
\begin{itemize}
\item
\Sergey{
%Four states of charged Higgs scalars are denoted as $\pi^\pm$ and $h^\pm$.
Four charged Higgs scalars, denoted by $\pi^\pm$ and $h^\pm$, with masses
% Their masses are
}
\begin{eqnarray}
m_{\pi^\pm}^2 &=& 0 \,,\\
m_{h^\pm}^2 &=& \frac{v_{H}^2+v_{\Sigma}^2}{2v_{H}v_{\Sigma}}(\sqrt{2} \kappa v_\Phi - \lambda'_{H\Sigma} v_H v_\Sigma ) \label{mhch}
\end{eqnarray}
The massless modes $\pi^\pm$ are \Sergey{ the charged would-be Nambu--Goldstone bosons of the spontaneously broken electroweak symmetry,
absorbed by
%which form
the massive $W^\pm$ bosons as their longitudinal components, while the mass eigenstates of the charged scalar fields $\pi^\pm$ and $h^\pm$ are linear combinations
%of the original fields:
\begin{eqnarray} \label{mixch}
\beginm{c}\pi^+ \\ h^+ \endm = \mathcal{U}_\mathrm{charged}
\beginm{c} a_{H}^+ \\ a_{\Sigma}^+ \endm  \,,
\end{eqnarray}
 of the original fields $a_{H}^+$ and $a_{\Sigma}^+$, where} $\mathcal{U}_\mathrm{charged}$ is the mixing matrix of the charged Higgs bosons shown in Eq.~\eqref{Ucharged}.

\item \Sergey{
Three neutral pseudo-scalars, denoted by $\pi^0$, $\eta^0$ and $a^0$, with masses:}
%Three states of neutral pseudo-scalar bosons are denoted by $\pi^0$, $\eta^0$ and $a^0$. }
%Their masses are
\begin{eqnarray}
m_{\pi^0}^2 &=& 0 \,,\\
m_{\eta^0}^2 &=& 0 \,,\\
m_{a^0}^2 &=& \frac{\kappa}{\sqrt{2}} \frac{v_{H}^2v_{\Sigma}^2+v_{\Sigma}^2v_{\Phi}^2+v_{\Phi}^2v_{H}^2}{v_{H}v_{\Sigma}v_{\Phi}}\,. \label{ma}
\end{eqnarray}
\Sergey{The massless mode $\pi^0$ is the neutral would-be Nambu--Goldstone boson of the spontaneously broken electroweak symmetry,
%which forms
absorbed by}
the massive $Z^0$ boson as its longitudinal component. The massless mode $\eta^0$ is the neutral Nambu--Goldstone boson of the spontaneously broken lepton number $U(1)_{L}$ symmetry, called \textit{Majoron}. The mass eigenstates of the pseudo-scalar fields are the linear combinations of the original fields
\begin{eqnarray}\label{mixa}
\beginm{c} \pi^0 \\ \eta^0 \\ a^0 \endm =
\mathcal{U}_\mathrm{pseudo}
\beginm{c} a_{H}^0 \\ a_{\Sigma}^0 \\ a_{\Phi}^0 \endm  \,,
\end{eqnarray}
where $\mathcal{U}_\mathrm{pseudo}$ is the mixing matrix of the pseudo-scalar Higgs bosons, shown in Eq.~\eqref{Upseudo}.

\item  \Sergey{Three neutral scalar bosons,  denoted by $h^0$, $H^0$ and $s^0$.
%Three states of neutral scalar bosons are denoted by $h^0$, $H^0$ and $s^0$.
%Their masses are
In the rest of this paper we assume the hierarchy
%If we accept the hierarchy
$v_\Sigma,v_\Phi\ll v_H,\kappa$, which we motivate in what follows. In this case  their masses are approximately}
\begin{eqnarray}
m_{h^0}^2 &\sim & \lambda_H v_{H}^2\,,\label{Mh}\\
m_{H^0}^2 &\sim & \frac{\kappa v_H}{\sqrt{2}}\frac{v_{\Sigma}^2+v_{\Phi}^2}{v_{\Sigma}v_{\Phi}}\,,\label{MH}\\
m_{s^0}^2 &\sim & \mathcal{O}(v_{\Sigma}^2,v_{\Sigma}v_\Phi,v_{\Phi}^2) \,.\label{Ms}
\end{eqnarray}
The mass eigenstates are again linear combinations of the original fields:
\begin{eqnarray}\label{mixhHs}
\beginm{c} h^0 \\ H^0 \\ s^0 \endm
%\sim
=
\mathcal{U}_{hHs}
\beginm{c} h_{H}^0 \\ h_{\Sigma}^0 \\ h_{\Phi}^0 \endm  \,,
\end{eqnarray}
where $\mathcal{U}_{hHs}$ is the mixing matrix of the pseudo-scalar Higgs bosons,
\Sergey{
%Its approximation is shown in Eq.~\eqref{UhHs}.
whose approximate form is given in Eq.~\eqref{UhHs}.

In this spectrum we identify $h^0$ with the SM Higgs boson. There are also a heavy Higgs boson, $H^0$, with a significant electroweak coupling,
and  a very light SM-sterile scalar $s^0$. We will specify the scalar Higgs boson spectrum with more details in the subsequent sections.
%better later, namely the estimation of $m_{s^0}^2$, after we learn about hierarchies among other model parameters so that we could make approximations in a more sophisticated way.
%
%From that we can identify $h^0$ as the SM-like Higgs boson, $H^0$ as an additional Higgs boson having significant electroweak interactions that should be better made heavy, and $s^0$ as an additional very light sterile scalar. We will specify the scalar Higgs boson spectrum better later, namely the estimation of $m_{s^0}^2$, after we learn about hierarchies among other model parameters so that we could make approximations in a more sophisticated way.
}
\end{itemize}

\subsection{The coupling constants of Higgs bosons}

Once we know the linear combinations of the original fields forming the mass eigenstates of Higgs bosons and neutrinos, we can derive expressions for neutrino Yukawa, gauge and other coupling constants. The coupling constants can be read from the Lagrangian $\eL_\mathrm{eff}$ in Eq.~\eqref{L} after the Higgs fields are shifted by their
%vacuum expectation values
VEVs,
according to Eq.~\eqref{shift}, and the Higgs and neutrino fields are
\Sergey{
replaced with the mass eigenstates, according to \eqref{mixch}, \eqref{mixa}, \eqref{mixhHs} and \eqref{mixnu}.
%
%substituted for the appropriate linear combinations of the mass eigenstates, according to \eqref{mixch}, \eqref{mixa}, \eqref{mixhHs} and \eqref{mixnu}.
}

\section{Low-energy solution estimates}\label{Pheno}

Now we have all ingredients to check the phenomenological viability of the model. First we solve the RGEs and then, in terms of low-energy values of the model parameters, we determine the particle masses and interactions. At this first stage of the model development we perform just order-of-magnitude estimates.
Surprisingly, we find that in our model
%although surprisingly, we find that in our model
there is small room
%for the model
to play with the parameters, in order to
%to satisfy
satisfy phenomenological and theoretical constraints.
%of the model.
The low-energy values of the parameters are constrained by the existing experimental data on particle masses and coupling constants  \cite{PhysRevD.98.030001},
%, which we want to reproduce.
%
%The low-energy values of the parameters are fixed by the values and constraints of the particle masses and coupling constants, which we want to reproduce.
%
while the high-energy values of the parameters are fixed by the boundary conditions
(\ref{Yuk_bound})-(\ref{kappa_bound}) at the scale $\Lambda$, where the effective model must be matched with the underlying four-fermion interactions. Admittedly, it might easily happen that the model 
%does not meet these constraints that would rule it out.
does not meet these constraints, in which case it would be ruled out.
%Clearly, it might easily happen that these constraints are not able to rule out the model.

In the following,  we first discuss the general features of the RGE solution and list the typical order of magnitude low-energy values of the dynamically generated coupling parameters $y_{\Sigma}$, $y_{\Phi}$, $\lambda_{\Sigma}$, $\lambda_{\Phi}$, $\lambda_{H\Sigma}$, $\lambda'_{H\Sigma}$, $\lambda_{\Phi H}$, $\lambda_{\Phi\Sigma}$ and $\kappa$.
\Sergey{
Thus, these are not really free parameters of the model, since they have been fixed from the RGEs with the corresponding  boundary condition (\ref{Yuk_bound})-(\ref{kappa_bound}).
%
%They are not free parameters of the model, they just result from the GR equations with their given boundary condition \eqref{Yuk_bound}, \eqref{lambda_bound} and \eqref{kappa_bound}.
It is important to point out that their low-energy values depend only weakly on the actual value of the compositeness scale $\Lambda$, which we fix by
%rather dubious
the requirement of one-loop vacuum stability. Next we fix a set of the SM-sector parameters, $v_H$, $y_t(m_t)$, and $\lambda_H$,
%by reproducing
from the experimental values of the masses of the electroweak gauge bosons, the SM Higgs and the top-quark. As a result we end up with only four seesaw-related free parameters
% which are
\begin{equation}\label{free_parameters}
	M_R,\ v_\Phi,\ v_\Sigma,\ y_H(M_R)\,.
\end{equation}
In what follows we will show how to limit them from the non-observation of extra Higgs bosons at the LHC  and from Leptogenesis.
%
%On them we apply constraints of additional Higgs bosons which need to be aligned with their non-observation at LHC and constraints motivated by Leptogenesis.
\footnote{Within our simplified single-flavor model,  $CP$ violation is missing, and thus Leptogenesis does not work. Still we adapt constraints from the realistic three-flavor model on the Yukawa coupling strengths and on the heavy neutrino mass splitting, and apply them to our model.}
%
%Finally, we discuss the model predictions for the light neutrino mass and for the properties of a dark % matter candidate.
}

\subsection{General features of the RGE solution}

\begin{figure}[h]
	\includegraphics[width=0.9\columnwidth]{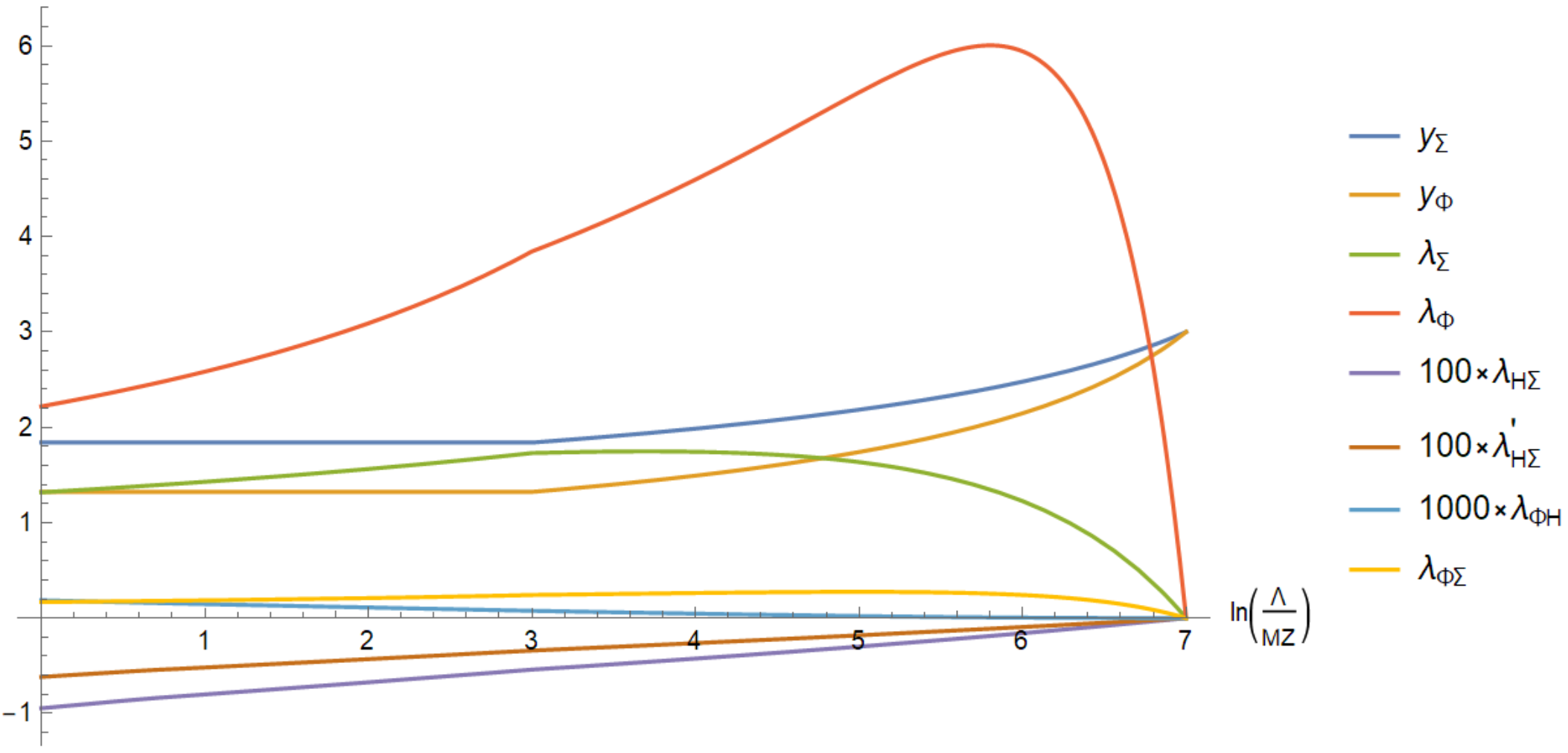}
	\caption{A typical solution of the RGEs for $y_0=3$, $\Lambda=\e^{7}M_Z\approx100\TeV$ and $M_R=\e^{3}M_Z\approx1.8\TeV$.}
	\label{RGE}
\end{figure}

\begin{figure}[h]
	\includegraphics[width=0.9\columnwidth]{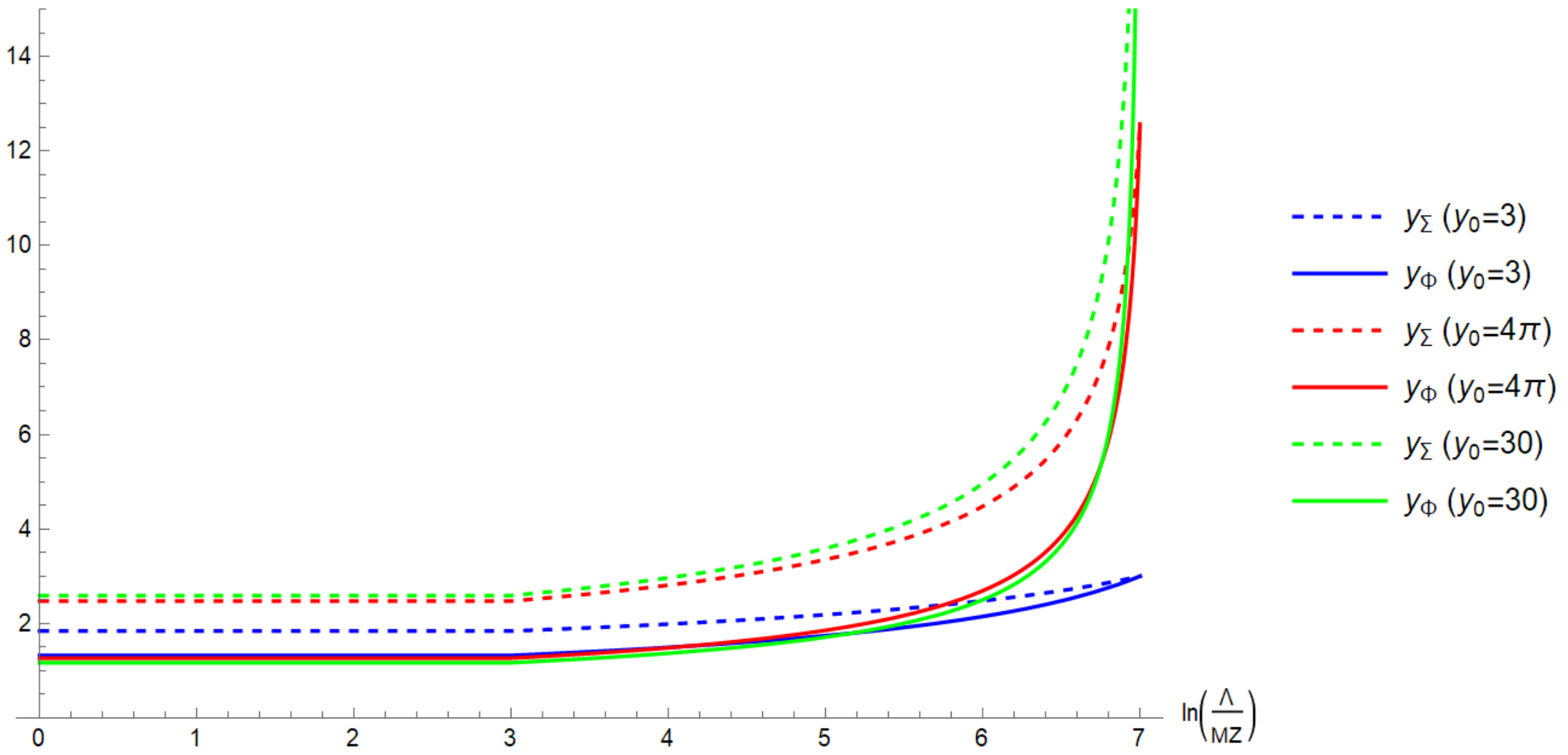}
	\caption{A solution for $y_\Sigma$ and $y_\Phi$ of the RGEs for the boundary conditions \eqref{Yuk_bound} given by three values $y_0=3,4\pi,30$, and for $\Lambda=\e^{7}M_Z\approx100\TeV$ and $M_R=\e^{3}M_Z\approx1.8\TeV$.}
	\label{y_fixpunkt}
\end{figure}

\Sergey{As we already stated,
the effective parameters $y_{\Sigma}$, $y_{\Phi}$, $\lambda_{\Sigma}$, $\lambda_{\Phi}$, $\lambda_{H\Sigma}$, $\lambda'_{H\Sigma}$, $\lambda_{\Phi H}$, $\lambda_{\Phi\Sigma}$ and $\kappa$ are not free model parameters, since their low-energy values are determined by the solution of the corresponding RGEs shown in Appendix \ref{AppRGE}, with the high-scale boundary conditions (\ref{Yuk_bound})-(\ref{kappa_bound}).
%Due to the dynamical origin of
%the effective parameters $y_{\Sigma}$, $y_{\Phi}$, $\lambda_{\Sigma}$, $\lambda_{\Phi}$, $\lambda_{H\Sigma}$, $\lambda'_{H\Sigma}$, $\lambda_{\Phi H}$, $\lambda_{\Phi\Sigma}$ and $\kappa$ their low-energy values are not free parameters. They just result from their RG equations.
}
A typical solution is plotted in the Fig.~\ref{RGE}.
The Yukawa parameters $y_\Sigma$ and $y_\Phi$, starting from their value $y_0$ given by the boundary condition \eqref{Yuk_bound} at the compositeness scale $\Lambda$, do not run to small $\ll1$ values. Typically, they are
\begin{eqnarray}
y_\Sigma(M_R) &\simeq& 1\,, \label{ySMR}\\
y_\Phi(M_R) &\simeq& 1\,. \label{yFMR}
\end{eqnarray}
Such estimate is in fact rather robust \cite{Bardeen:1989ds,Cvetic:1997eb}, since it is sensitive only very weakly to the high-energy values of $y_\Sigma$ and $y_\Phi$, and exhibiting the typical behavior in the presence of an infrared fixed point, which is demonstrated in the Fig.~\ref{y_fixpunkt}. We can see that a rather wide range of high-energy Yukawa parameter values, $y_0\in(3,30)$, is squeezed by RGE evolution into a quite small range of low-energy values, $y_\Sigma(M_R)\in(1.8,2.6)$ and $y_\Phi(M_R)\in(1.1,1.4)$.\footnote{Notice that due to the $M_R$ threshold, the RGE evolution of the Yukawa parameters takes place only within the interval $(M_R,\Lambda)$, while below $M_R$ they
%Yukawa parameters
freeze.}

On the other hand, the neutrino Yukawa coupling $y_H$ is not generated dynamically, and therefore, it is not subject to any boundary condition at $\Lambda$. Thus, its value, $y_H(M_R)$, is a free parameter to be fixed from phenomenology. In particular, we will fix it  later by arguments of successful leptogenesis.
% \\
%\textcolor{red}{What about limits from invisible Higgs decay?}

All the scalar quartic couplings $\lambda$'s, except for $\lambda_H$, are generated dynamically,
%That is expressed as a strict high-energy boundary condition \eqref{lambda_bound}.
and fixed by the high-energy boundary conditions \eqref{lambda_bound}.
To determine the mass spectrum of the Higgs bosons we need to know the low-energy values of these quartic couplings, which are solutions of the corresponding RGEs given in Appendix \ref{AppRGE}.
Our analysis of their solutions in a wide range of the free model parameters shows that the quartic couplings typically demonstrate the following hierarchy

%As a matter of solving the system of RG equations and depending on the particular choice of the free parameters, we typically get the following hierarchy of the quartic couplings
%
\begin{eqnarray}
\lambda_\Sigma,\lambda_\Phi (M_Z) & \sim & \ \mathcal{O}(1) \,,\label{lambdaPhi}\\
\lambda_{\Phi\Sigma} (M_Z)& \sim & \ \mathcal{O}(10^{-1}) \,,\\
\lambda_{H\Sigma},\lambda'_{H\Sigma} (M_Z)& \sim & -\mathcal{O}(10^{-2}) \,,\label{lambdaHS}\\
\lambda_{\Phi H} (M_Z)& \sim & \ \mathcal{O}(10^{-4}) \label{lambdaFH}\,.
\end{eqnarray}

The $\kappa$ parameter is also generated dynamically.
It is fixed by the high-energy boundary condition \eqref{kappa_bound}, and
%It is expressed as the strict high-energy boundary condition \eqref{kappa_bound}.
its magnitude is driven mainly by the last term in Eq.~\eqref{RGEkappa}, as can be seen from
\begin{eqnarray}
%:{SK 15.02.2019-1} I eliminated the prime sign here. I guess it was a typo.
%\D\kappa' & = &
\Sergey{\D\kappa} & = &
\kappa f - 8\, y_{H}\, y_{\Sigma}\, y_{\Phi}\, M_R \,,
\end{eqnarray}
where
\begin{equation}
f= 2\lambda_{\Phi H} + 2\lambda_{\Phi\Sigma} + 2\lambda_{H\Sigma} + 4\lambda'_{H\Sigma} - \tfrac{3}{2}\big(3g_{2}^{2}+g_{1}^{2}\big) + y_{H}^2 + y_{\Sigma}^2 + 2y_{\Phi}^2 + 3 y_{t}^2 \,.
\end{equation}
Since the last term drops off below the heavy neutrino decoupling scale $M_R$, the evolution of
$\kappa$
\Sergey{saturates at this point.
%saturates there.
Therefore, in order to estimate the magnitude of $\kappa$ it is sufficient to calculate its value at $M_R$. Neglecting the scale dependence of all the other parameters, we can write
%the solution at $M_R$ as
}
\begin{equation}\label{kappa_est}
%:{SK 15.02.2019-2} I changed 32 --> 16. Please, check.
\kappa(M_R) \approx \frac{8\, y_{H}\, y_{\Sigma}\, y_{\Phi}}{f}
%\left[1-\left(\frac{M_R}{\Lambda}\right)^{f/32\pi^2}\right]M_R\  \sim y_H M_R \,.
\left[1-\left(\frac{M_R}{\Lambda}\right)^{f/16\pi^2}\right]M_R\  \sim y_H M_R \,.
\end{equation}
\Sergey{

The last very rough estimate is obtained from our numerical analysis of the low-energy values of the model Yukawa, gauge and quartic coupling constants. It follows that typically  $f\sim\mathcal{O}(10)$, so the exponent in the second term
%
%The last very rough estimate follows from $f\sim\mathcal{O}(10)$, so the exponent in the second term
is $<1$, but not $\ll1$. Then taking into account (\ref{nondecoupling}) we neglect the second term in the square bracket.
%less than unity, but not much smaller.
%smaller than unity, but not much smaller.
%With the hierarchy among $M_R$ and $\Lambda$, the square braket gives a factor of $\mathcal{O}(1)$.
Since the ratio $8\, y_{\Sigma}\, y_{\Phi}/f$ is of order $\mathcal{O}(1)$, we come up with the above-mentioned  rough estimate.}

\subsection{Fixing of parameters related to the SM}

The Higgs-doublet VEVs must satisfy
\begin{equation}
v=\sqrt{v_{H}^2+v_{\Sigma}^2}\doteq246\GeV \,,
\label{newvev}
\end{equation}
in order to get the correct values for the masses of $W$ and $Z$. To achieve the hierarchy, from Eq.~\eqref{seesaw_hierarchy} or \eqref{seesaw_hierarchyx} and taking into account the value of the Yukawa parameter $y_\Sigma$ in Eq.~\eqref{ySMR}, the hierarchy of the VEVs $v_\Sigma\ll v_H$ is required. From that we can set
\begin{equation}
v_{H}\simeq v\doteq246\GeV \,.
\label{v_approx}
\end{equation}
Based on this, in order to reproduce the mass of the top quark as
\begin{equation}
m_t=y_t(m_t) \, v_H/\sqrt{2}\doteq174\GeV .
\end{equation}
the fixing of the top-quark Yukawa parameter as
\begin{equation}
y_t(m_t)\approx \frac{\sqrt{2}m_t}{v_H}\doteq1 \,,
\end{equation}
is required, just as in the SM.

The quartic coupling parameter of the elementary Higgs boson, $\lambda_H$, is fixed by the phenomenological requirement to reproduce the SM-like Higgs boson mass, $m_{h^0}\doteq 125\GeV$. The mass of the SM-like Higgs boson in our model is approximately given by Eq.~\eqref{Mh}, which leads to
\begin{equation}\label{lambdaHfix}
\lambda_H(m_{h^0})\simeq\frac{m_{h^0}^2}{v_{H}^2}\doteq 0.26 \,.
\end{equation}
Therefore, we should set the initial value $\lambda_H(\Lambda)$ in a way that $\lambda_H(m_{h^0})$ gets the value stated in Eq.~\eqref{lambdaHfix}. On the other hand $\lambda_H(\Lambda)$ should not be negative, since
\Sergey{otherwise the ground state of the model would be unstable.
%That would otherwise mean that the elementary Higgs boson vacuum expectation value would be unstable, heralding some new physics below the scale $\Lambda$, i.e., our model would be not correct.
}
The one-loop RGEs show that the small value of $\lambda_H(m_{h^0})$ presented in \eqref{lambdaHfix}, requires that the initial value $\lambda_H(\Lambda)$ be negative, unless
\begin{equation}\label{vacstab}
\Lambda\leq 100\TeV \,.
\end{equation}
This feature of our model is practically the same as in the SM, where the one-loop RGEs exhibit the same limit on the vacuum stability.
\Sergey{
Nevertheless, the three-loop level RGEs of the SM show \cite{Degrassi:2012ry} that the vacuum stability limit is, in fact, pushed to much higher scales.
%
%and, therefore, the limit \eqref{vacstab} should not be taken too
%strictly.
%seriously.
}
Although we expect the same behavior in our model, to be on the safe side we set the value of the compositeness scale to be
\begin{equation}\label{Lambda_fix}
\Lambda=100\TeV \,.
\end{equation}
\Sergey{
%This means that we do not take $\Lambda$ as a free parameter, even though, strictly speaking, we should. Our experience with
%the numerical solution is however such,
Notice that the RGE solutions are weakly sensitive to the actual value of
$\Lambda$. }

\subsection{Constraint from additional Higgs bosons}
\label{sec:ExtraHiggsConstr-1}
\Sergey{As we already saw that our model contains extra Higgses. In order to pass the existing experimental constraints  \cite{PhysRevD.98.030001}
%On the same footing as two-Higgs-doublet models, for instance, our model predicts Higgs particles additional to the SM one. As such it it is subject of constraints in the current times when the LHC is seeing nothing else but the Standard Model.
%The additional Higgs particles
they
must be either sufficiently heavy, i.e., with masses greater than $500\GeV$, or sufficiently weakly coupled to the known SM particles, i.e., the light states should be predominantly made of sterile
$\Phi$ fields with only a small admixtures of electroweak doublets $H$ and $\Sigma$ fields. From these constraints, a large ratio
}
\begin{equation}\label{rFSgg1}
\frac{v_\Phi}{v_\Sigma}\equiv r_{\Phi\Sigma} \gg 1
\end{equation}
can be advocated as follows.
%
%\Adam{The vev $v_\Phi$ will be the subject of the hierarchy $v_\Phi\ll v_H$ as well. As the phenomenology will require, the hierarchy off all three masses will be
%\begin{equation}\label{vHvSvF}
%v_\Sigma\ll v_\Phi\ll v_H\,.
%\end{equation}
%Throughout our analysis, $v_\Sigma$ and $v_\Phi$ are free parameters. We will sometimes use the ratio
%\begin{equation}\label{rFS}
%r_{\Phi\Sigma}\equiv\frac{v_\Phi}{v_\Sigma} \,.
%\end{equation} }

Assuming the hierarchy $v_\Sigma,v_\Phi\ll v_H,\kappa$ and using the estimate \eqref{kappa_est} in Eq.~\eqref{mhch}, we get an expression for the charged Higgs boson mass
\footnote{In the following, we do not indicate explicitly
%will suppress the argument of
the RGE scale $m$ dependence of the running coupling constants.
% while keeping in mind
It is implicit that the Yukawa coupling constants
%should be
are evaluated at $m=M_R$, while the others
% coupling constants should be evaluated
at the mass of the corresponding particle.}
%
%\footnote{In the following, we will suppress the argument of the scale $m$ of the running coupling constants while keeping in mind that the Yukawa coupling constants should be evaluated at $m=M_R$ while the others
%% coupling constants should be evaluated
%at the mass of the corresponding particle.}
%
\begin{eqnarray}
m_{h^\pm}^2 &\approx& \frac{v_H}{2}\Big(\sqrt{2} \, r_{\Phi\Sigma} \, M_R \, y_H-\lambda'_{H\Sigma}\, v_H\Big) \,.
\label{chargedhmass}
\end{eqnarray}
If the second term is larger in magnitude than the first one, we would have the problem of having a too light charged Higgs boson with mass $\sim 0.1 v$,
%from
\Sergey{according to the estimate in
Eq.~\eqref{lambdaHS}.
The first term contains the product $M_R \, y_H$, which in our seesaw scenario
easily turns out to be of
%is easily of
%can be easily set to
the same order of magnitude as $v$ or even smaller. Therefore, to make the first term dominant in Eq.~\eqref{chargedhmass},  we need $r_{\Phi\Sigma}$ to be large enough, as indicated in Eq.~\eqref{rFSgg1}.

Among the pseudo-scalars there is the majoron $\eta^0$, which is a massless Nambu--Goldstone boson of the spontaneously broken $U_{L}(1)$. In order to
make it phenomenologically harmless,
%it must be predominantly a
we require that it should be dominated by the SM singlet
$a_{\Phi}^0$ state. From Eq.~\eqref{mixa} and Eq.~\eqref{Upseudo} we see that the majoron $\eta^0$
%lacks
does not contain
%the admixture of
$a_{H}^0$ component, and
therefore,
%and that
we only need to suppress the $a_{\Sigma}^0$ admixture,
%we need
requiring again the condition
%$r_{\Phi\Sigma}$ again large, i.e.,
given in Eq.~\eqref{rFSgg1}.

However, the massive pseudo-scalar $a^0$ cannot be made sterile, hence we must guarantee it to be  sufficiently heavy in order to pass the experimental constraints from the neutral Higgs non-observation  \cite{PhysRevD.98.030001}.
%
%The massive pseudo-scalar $a^0$ is however not possible to make sterile, hence we need to make it sufficiently heavy.
}
The approximate value of its mass, under the assumption of the hierarchy $v_\Sigma,v_\Phi\ll v_H,\kappa$, is
\begin{equation}\label{ma_est}
m_{a^0}^2 \approx \left(r_{\Phi\Sigma}+\frac{1}{r_{\Phi\Sigma}}\right)\frac{y_H}{\sqrt{2}} v_H M_R  \,.
\end{equation}
Therefore, to make $a^0$ heavy we have now two possibilities for $r_{\Phi\Sigma}$, either very large or very small, and then to be compatible with the previous requirements we are again led to Eq.~\eqref{rFSgg1}.

\Sergey{
A similar situation takes place in the sector of  the $H^0$ and $s^0$  Higgs bosons. From
%their masses,
Eq.~\eqref{MH} and Eq.~\eqref{Ms}, using Eq.~(\ref{kappa_est}) we have that their masses are
%
%Similar situation appears among the additional two scalar Higgs bosons $H^0$ and $s^0$. From their masses, \eqref{MH} and \eqref{Ms}, we see that
}
\begin{equation}\label{mH_est}
m_{H^0}^2 \approx \left(r_{\Phi\Sigma}+\frac{1}{r_{\Phi\Sigma}}\right)\frac{y_H}{\sqrt{2}} v_H M_R  \,.
\end{equation}
and
%that we cannot avoid the fact that
\begin{equation} \label{ms_est}
m_{s^0}\ll m_{h^0} \,.
\end{equation}
\Sergey{
Analogously to Eq.~(\ref{ma_est}), a large value of $r_{\Phi\Sigma}$ allows making $H^{0}$ sufficiently heavy, in accordance with the current experimental constraints  \cite{PhysRevD.98.030001}.
%one of the two extreme values, either very big or very small.
%
%The estimate \eqref{mH_est} again pushes $r_{\Phi\Sigma}$ to one of the two extreme values, either very big or very small.
%
However, the scalar $s^0$ is unavoidably light (\ref{ms_est}) and, therefore, must be
%made
%sufficiently
predominantly
sterile.
As follows from
Eqs.~\eqref{mixhHs} and \eqref{UhHs}, this condition is satisfied for
%to choose the option of
large values of $r_{\Phi\Sigma}$ \eqref{rFSgg1}, and then,
% again.
%
under the hierarchy $v_\Sigma\ll v_\Phi\ll v_H$ and $r_{\Phi\Sigma}\gg1$ and $\lambda_{\Phi H}\ll\lambda_H$,
%we can specify better
the $s^0$-boson mass can be approximated by
\begin{equation}
m_{s^0}^2 \approx \lambda_\Phi v_{\Phi}^2\,.
\end{equation}
%and the scalar mixing matrix $\mathcal{U}_{hHs}$ as \eqref{UhHs}.

Let us summarize this section.
In order to satisfy the phenomenological requirements it is necessary to consider large values of $r_{\Phi\Sigma}$ \eqref{rFSgg1}.
%
%In this  from which it follows that
Then
we can identify two groups of Higgs bosons: }
\begin{itemize}
	\item \Sergey{Light Higgs bosons: the SM-like Higgs boson $h^0$, made mostly of the elementary electroweak doublet $H$ field; a very light scalar $s^0$ and massless majoron $\eta^0$, both made mostly of the singlet $\Phi$. Their masses
%below
at the electroweak scale are
	\begin{eqnarray}
	m_{h^0}^2 & \approx & \lambda_H v_{H}^2\,,\\
	m_{s^0}^2 & \approx & \lambda_\Phi v_{\Phi}^2\,,\label{ms}\\
	m_{\eta^0}^2 & = & 0\,,
	\end{eqnarray}
}
	\item \Sergey{Heavy Higgs bosons: $X=h^\pm,\,a^0,\,H^0$, which are all made mostly of the electroweak doublet $\Sigma$
%, with their
and have almost degenerate masses above the electroweak scale:
	\begin{eqnarray}
	m_{X}^2 & \approx & r_{\Phi\Sigma} m_D M_R\,.\label{mX}
	\end{eqnarray}}
\end{itemize}
%To complete the list, we should mention three SM-like would-be Nambu--Goldstone bosons
%$\pi^\pm$ and $\pi^0$, made mostly of the Higgs doublet $H$.
%

\subsection{Constraints from Leptogenesis}

\Sergey{
Leptogenesis renders stringent
%phenomenological tool to
constraints on the free parameters \eqref{free_parameters} of our model.
%
%In the case of our model the Leptogenesis plays role of the most powerful phenomenological tool to constrain the free parameters, namely those in \eqref{free_parameters}.
%
The Sakharov conditions for the case of leptogenesis are: 1) LNV processes are allowed, 2) they are $CP$ asymmetric,
%($\cancel{CP}$),
3) before they become cosmologically irrelevant during the evolution of the Universe they go out of thermal equilibrium.
%
%The Sakharov conditions are, in the case of leptogenesis, translated into the following conditions: 1) The LNV processes are allowed, 2) they are $CP$ asymmetric, 3) before they become cosmologically irrelevant during the evolution of the Universe they go out of thermal equilibrium.
%
Obviously, the condition 1) is satisfied in our model. As to the condition 2), in the simplified version of our model with only one fermion generation, which we are studying here, $CP$ is conserved. However, $CP$ violation (CPV) can be easily accommodated in the \Adam{realistic} models of any seesaw scenario, since there are several leptonic Yukawa coupling constants in the neutrino mass matrix which are complex, leading to the physical CPV phases. However, the resulting CPV effect may not be sufficiently strong for successful leptogenesis.
%
%By definition of our model, the first condition about the presence of the LNV is clearly satisfied. As for the second condition, the $CP$ violation can in principle be easily accommodated in the models of any seesaw scenario as there is enough leptonic Yukawa coupling parameters forming the neutrino mass matrix that can be made complex. However the resulting $CP$ asymmetry may not be great enough.
%
In order to enhance the CPV one has to either push the masses of the heavy Majorana neutrinos \eqref{mN} to very large values $>10^9\GeV$ \cite{Buchmuller:1996pa}, or introduce a pair of quasi-degenerate heavy Majorana neutrinos, leading to resonant enhancement of the CPV in their LNV decays \cite{Pilaftsis:1997jf}. The first option is not pertinent for our model, due to Eqs.~\eqref{nondecoupling} and \eqref{vacstab}. Meanwhile, the second option is naturally realizable in our model, as in any other model with a low-scale seesaw.
The resonant condition providing the maximal CPV effect
%
%In order to obtain a sizable $CP$ asymmetry, either the mass of the heavy Majorana neutrinos \eqref{mN} should be very big $>10^9\GeV$ \cite{Buchmuller:1996pa}, which is due to \eqref{nondecoupling} and \eqref{vacstab} not the case in our model, or the model should rely on a resonant enhancement of the $CP$ asymmetry between the LNV decays of a pair of quasi-degenerate heavy Majorana neutrinos introduced in \cite{Pilaftsis:1997jf}, which the low-scale seesaw mechanism used in our model is designed for. The resonant condition for obtaining the maximal $CP$ asymmetry
%
\begin{equation}\label{2ndSakhar}
	\frac{\Gamma_N}{2}\simeq|m_{N_+}-m_{N_-}|
\end{equation}
relates the mass splitting of the heavy Majorana neutrinos and their decay rate $\Gamma_N$.
%tights together the mass splitting of the heavy Majorana neutrinos and their decay rate $\Gamma_N$.

Condition 3) requires that the expansion rate of the Universe, quantified by the temperature-dependent Hubble parameter $H(T)$, is larger than the decay rate of the heavy Majorana neutrinos,
\begin{equation}\label{3rdSakhar}
	H(T=m_{N_+}) \gtrsim\frac{\Gamma_N}{2}\,.
\end{equation}
The Hubble parameter at a temperature $T$ for a given extension of the SM with  $g_*$ degrees of freedom is $H(T)\sim1.73\sqrt{g_*}T^2/\Lambda_\mathrm{Planck}$.

%Our model contains a set of additional Higgs bosons, which allow Majorana neutrinos to decay through several channels into light neutrinos, which are important for leptogenesis.

%
Our model contains a set of extra Higgs bosons, providing heavy Majorana neutrinos with several decay modes into the light neutrino relevant for leptogenesis. The total decay rate of the heavy Majorana neutrinos is given by
%
%Our model contains a set of additional Higgs bosons our heavy Majorana neutrinos have several decay channels into a light neutrino relevant for leptogenesis. Their total decay rate is given as
\begin{equation}
	\Gamma_N\sim\sum_i\frac{y_{N\rightarrow i}^2}{8\pi}m_{N_+}\,,
\end{equation}
where $y_{N\rightarrow i}$ is a coupling constant responsible for the $i$-th decay channel. They can be  sorted into groups according to the boson emitted in the decay.
Here we present  approximate expressions for these couplings, derived with the assumption of the hierarchy $v_\Sigma\ll v_\Phi\ll v_H$ motivated in the previous sections.
\begin{itemize}
	\item Light Higgs boson emission\footnote{Here we have again suppressed the scale dependence of the running coupling constants.}
	\begin{eqnarray}
	y_{N\rightarrow\nu h} & \approx & -\frac{y_H}{2}\,,\label{yNnuh}\\
	y_{N\rightarrow\nu s} & \approx & \frac{1}{2r_{\Phi\Sigma}}y_\Sigma-\frac{m_D}{M_R}y_\Phi\,,\label{yNnus}\\
	y_{N\rightarrow\nu \eta} & \approx & \frac{1}{2r_{\Phi\Sigma}}y_\Sigma-\frac{m_D}{M_R}y_\Phi\,,\label{yNnueta}\\
	y_{N\rightarrow e h} & \approx & y_{H}^2\frac{v_\Sigma}{2M_R}\,.\label{yNeh}
	\end{eqnarray}
	\item Heavy Higgs boson $X=a^0,\,H^0$ emission
	\begin{eqnarray}
	y_{N\rightarrow\nu X} & \approx & -\frac{y_\Sigma}{2}\,.\label{yNnuX}
	\end{eqnarray}
%\end{itemize}
%
%For completeness, we list also the coupling constants for the other decay channels of the heavy neutrino into the gauge bosons and charged leptons:
\item The SM Gauge boson emission
\begin{eqnarray}
y_{N\rightarrow\nu Z} & \approx & \frac{m_D}{\sqrt{2}M_R}\frac{g}{2\cos{\theta_W}}\,,\label{gNnuZ}\\
y_{N\rightarrow e W} & \approx & \frac{m_D}{\sqrt{2}M_R}\frac{g}{\sqrt{2}}\,,\label{gNeW}
%\\
%y_{N\rightarrow e h} & \approx & y_{H}^2\frac{v_\Sigma}{2M_R}\,.\label{yNeh}
\end{eqnarray}
\end{itemize}
}

%------- *** --------
In most of the model realizations of the leptogenesis with resonant $CP$ asymmetry enhancement, it is not necessary to tune the model parameters exactly to the resonance given by \eqref{2ndSakhar}. It is enough to be in the vicinity of the resonance. According to the numerical analysis in \cite{Pilaftsis:1997jf} performed for $m_{N_\pm}=10\TeV$ and $|y_{N}|\sim10^{-6}$, the mass splitting of the heavy neutrinos
\begin{equation}\label{masssplit}
\frac{|m_{N_+}-m_{N_-}|}{m_{N_+}}\sim 10^{-7}\,
\end{equation}
is sufficient. In a more recent analysis in \cite{Orikasa2011}, where a wash-out effect mediated by $Z'$ with mass $\sim M_N$ is taken into account, a stronger $CP$ asymmetry is needed requiring several orders of magnitude smaller mass splitting than in Eq.~\eqref{masssplit}. In our scheme similar wash-out effects can be expected due to the interaction channels mediated by the heavy Higgs bosons with mass $\sim M_N$. However to reliably address the wash-out effects requires a detailed analysis within a realistic version of our model, which we leave for future work. For now we take \eqref{masssplit} as a numerical input to our analysis, keeping in mind that if needed the mass splitting can be made correspondingly smaller by pushing $v_\Phi$ to lower values.
%The CPV for successful leptogenesis should be at least at the level $10^{-7}-10^{-6}$ for heavy neutrino masses $m_{N_\pm}\gtrsim1\TeV$ \cite{Pilaftsis:1997jf}. This means that it is not necessary to tune the model exactly to the resonance, given by \eqref{2ndSakhar}, but it is still necessary to be in its vicinity. According to the numerical analysis in \cite{Pilaftsis:1997jf}, performed for $m_{N_\pm}=10\TeV$ and $|y_{N}|\sim10^{-6}$, the mass splitting of the heavy neutrinos
%\begin{equation}\label{masssplit}
%\frac{|m_{N_+}-m_{N_-}|}{m_{N_+}}\sim 10^{-7}\,
%\end{equation}
%is sufficient, so we take \eqref{masssplit} as a numerical input to our analysis.
%
%Adam 18.02.2019: I think that it should be the number of degrees of freedom of the effective model below \Lambda which that is relevant for leptogenesis. It is possible to calculate the number of d.o.f. of our simplified single-generation model, it is 68:
%68 = 30(SM fermion d.o.f. in one generation) + 4(\nu_R+S_R d.o.f.) + 24(gauge bosons d.o.f.) + 10(2doublets+1singlet higgs d.o.f.)
%But the problem is, that we should better operate here with the number of d.o.f. corresponding to the realistic 3-generations model, which we have not formulated yet. And I do not know how many composite scalars and how many RH neutrinos we will need to operate with. But the estimate can be done by just multiplying the # of fermion d.o.f. by 3:
%136 = 3x(30 + 4) + 24 + 10
%

\Sergey{Estimating approximately the number of degrees of freedom in our model for the realistic 3-generation case  to be
$g_*\approx 100$
%
%=  90(SM fermion DOF) + $4\times 3$ ($\nu_R+S_R$ DOF) + 24(gauge bosons DOF) + 10 (2 doublets+1 singlet Higgs DOF) = 136.
%Assuming $g_*\sim100$ for our model, from \eqref{3rdSakhar},
}
we obtain an upper bound for the coupling constants
\begin{equation}\label{outofequilibrium}
y_{N\rightarrow i}\lesssim 10^{-7}\left[\frac{m_{N_+}}{\TeV}\right]\,.
\end{equation}

%Of course, our simplified model with one generation is not able to accommodate the leptogenesis, as all coupling parameters are real numbers and no CPV is present. However, we still use the leptogenesis as a guiding tool and apply the constraints \eqref{masssplit} and \eqref{outofequilibrium} to fix the magnitudes of the model parameters as we believe that these will not change dramatically when the model is reformulated in its full realistic three-generation version where the $CP$ violation is present.

The right-handed neutrino mass parameter $M_R$ is among the free parameters, ranging in our model roughly from $\sim1\TeV>v$ up to $\lesssim100\TeV=\Lambda$.
\Sergey{
In order to be specific let us consider a benchmark scenario in our model with
%In order to be specific and to present here a benchmark scenario of our model we fix $M_R$ to the value
}
\begin{equation}\label{MRfix}
M_R= 10\TeV\,.
\end{equation}
which enable us to relate directly our analysis to the conclusions of Ref.~\cite{Pilaftsis:1997jf}.

Now, all the coupling constants (\ref{yNnuh})-(\ref{gNeW}) must satisfy the out-of-equilibrium condition \eqref{outofequilibrium}.
\Sergey{
Therefore, the Yukawa parameter $y_H$ at the scale $M_R$ must be set at such a small value that the coupling constant \eqref{yNnuh} satisfies \eqref{outofequilibrium}, so we choose
%Therefore, the Yukawa parameter $y_H$ at the scale $M_R$ must be set sufficiently small so that the coupling constant \eqref{yNnuh} is small enough. We fix it as
}
\begin{equation}
y_H(M_R)\approx 10^{-7}\,,
\end{equation}
and with this value we calculate from \eqref{MRfix} and \eqref{mD} the ratio
\begin{equation}\label{mDMR}
\frac{m_D}{M_R}\approx 10^{-9}\,.
\end{equation}

Next we observe that the Yukawa coupling constants to heavy neutrinos \eqref{yNnuX} are unavoidably large,  a consequence of the dynamical origin of the Yukawa parameters $y_\Sigma$ and $y_\Phi$, which are of the order $\mathcal{O}(1)$ as a result of the RGE running from their large value at the compositeness scale $\Lambda$.
\Sergey{
Therefore, the only possibility to prevent the decay rate of  the heavy neutrinos from being unbearably large is to forbid the decay channels to heavy Higgs bosons kinematically by the condition $m_X>m_N$. Using the expression for the masses of heavy Higgs bosons \eqref{mX} and for heavy neutrinos $m_N\sim M_R$, we obtain the condition
%
%The only chance to prevent the decay rate of heavy neutrinos from being unbearably large is to forbid the decay channels to heavy Higgs bosons kinematically by the condition $m_X>m_N$. By using the expression for masses of heavy Higgs bosons \eqref{mX} and for heavy neutrinos $m_N\sim M_R$, we obtain the condition
%
\begin{equation}
\label{eq:r-MR}
r_{\Phi\Sigma}>\frac{M_R}{m_D}\,.
\end{equation}
Taking the ratio \eqref{mDMR} into account, we estimate $r_{\Phi\Sigma}$ conservatively as
%not to be an order of magnitude larger
\begin{equation}\label{rFSest}
r_{\Phi\Sigma}\approx 10^{9}\,.
\end{equation}
Now, in order to obtain the necessary CPV magnitude for successful leptogenesis, the mass splitting of the quasi-degenerate heavy neutrinos has to be at least of the order $10^{-7}$, see \eqref{masssplit}.
%Now, to obtain sufficiently large CPV the mass splitting of the quasi-degenerate heavy neutrinos has to be at least of the order $10^{-7}$, see \eqref{masssplit}.
}
From \eqref{mN} we see that the mass splitting is dominated by the inverse-seesaw mass parameter $\mu_\mathrm{inv}$
\begin{equation}
|m_{N_+}-m_{N_-}|\approx \mu_\mathrm{inv} = \frac{y_\Phi v_\Phi}{\sqrt{2}}\,.
\end{equation}
Provided that $y_\Phi\sim 1$, the mass splitting constrains the VEV of the SM singlet scalar $\Phi$ to
\begin{equation}
v_\Phi\le 10^{-7}M_R\,.
\end{equation}
%:{SK 17.02.2019-3} Maybe we should choose this value from the beginning in (102)?
\Sergey{
To be more safe with leptogenesis we may choose in Eq.~(\ref{masssplit}) the mass splitting to be $10^{-8}$, which leads to
%Let us be generous and set the mass splitting to $10^{-8}$ which means
}
\begin{equation}\label{vPhi}
v_\Phi\approx 100\keV\,.
\end{equation}
From \eqref{rFSest} we obtain
\begin{equation}
v_\Sigma\approx 0.1\meV\,.
\end{equation}
\Sergey{
%That completes the order of magnitude estimate of the model free parameters and in what follows we will discuss  several predictions of our model.
%
%That completes the fixing of the free parameters and in the rest of this subsection and in the following subsections we can see what are the consequences.

Let us summarize our order-of-magnitude estimates of the couplings constants
(\ref{yNnuh})-(\ref{gNeW}), motivated by successful  leptogenesis:
%of light bosons \eqref{yNnus} and \eqref{yNeh} should also be of the order $10^{-6}$ at most.
%Using the fixed values of the free parameters we get totally safe situation
%
%In order to complete the check of the leptogenesis, the rest of the Yukawa couplings of light bosons \eqref{yNnus} and \eqref{yNeh} should also be of the order $10^{-6}$ at most.
%Using the fixed values of the free parameters we get totally safe situation
}
\begin{eqnarray}
y_{N\nu s} & \approx & 10^{-9}\,,\\
y_{N\nu \eta} & \approx & 10^{-9}\,,\\
y_{N\nu Z} & \approx & 10^{-9}\frac{g}{2\cos{\theta_W}}\,,\\
y_{N e W} & \approx & 10^{-9}\frac{g}{\sqrt{2}}\,,\\
y_{N e h} & \approx & 10^{-31}\,.
\end{eqnarray}
\Sergey{This completes the estimation of the free parameters of our model and in what follows we will discuss  some of its predictions.}

\section{Prediction of the model}

\Sergey{With the parameters of our model approximately evaluated in the previous sections, we can derive its key predictions and compare with the existing experimental data. As will be seen, there is a small room to play with the parameters within the ballpark of the approximations made in these evaluations. Therefore, the model is predictive and falsifiable. }
%Up to this point of our analysis, we have fixed all parameters of the model. Still there are phenomena to be reproduced or predicted. That is an expression of the robustness of our model, that there is a small room to play with parameters. The model is easily falsifiable.

\subsection{Light neutrino mass}
\label{sec:Light neutrino mass}

\Sergey{
The tiny active neutrino mass is given by Eq.~\eqref{mnuw} in terms of the Weinberg parameters, $w_\mathrm{inv}, w_\mathrm{lin}$, rather than the Yukawa couplings. Since the RGE running of these Weinberg parameters is quite moderate, their order of magnitude stays the same over large interval of scales, from $M_R$ down to $m_\nu$. Therefore, in order to estimate the neutrino mass it is sufficient to consider just the initial values of the Weinberg parameters at the scale $M_R$, given in Eqs.~\eqref{wyHyF} and \eqref{wyHyS}.
%, i.e.,
%\begin{eqnarray}
%w_\mathrm{inv} &\approx & \left. y_{H}^2y_\Phi\right|_{m=M_R} \,, \\
%w_\mathrm{lin} &\approx & \left. -2y_{H}y_\Sigma\right|_{m=M_R} \,.
%\end{eqnarray}
Inserting these values into \eqref{mnuw} and applying \eqref{mD} we obtain
%the expression, which under the order-of-magnitude fixing of all parameters from the previous section \ref{Pheno}, gives the active neutrino incredibly light
\begin{equation}
\label{eq:light-nu-mass}
m_\nu \approx \frac{v_\Phi}{\sqrt{2}}\frac{m_D}{M_{R}} \left(\frac{m_D}{M_{R}}y_\Phi - \frac{2}{r_{\Phi\Sigma}}y_\Sigma \right)
\lesssim 10^{-13}\eV  \, ,
%\approx 10^{-13}\eV \, ,
\end{equation}
where in the numerical estimation we used (\ref{mDMR}), (\ref{rFSest}) and (\ref{vPhi}).
%the expression, which under the order-of-magnitude fixing of all parameters from the previous section \ref{Pheno}, gives the active neutrino incredibly light
%
Thus, the model predicts an extremely light active neutrino. It is  important to note that this prediction can hardly be avoided in the present single-generation version of the  model.
%
%This is a prediction of our model which can hardly be avoided in the present single-generation model. In the case of the realistic three-generation model we can expect similar result.
In the case of the realistic three-generation version of the model we expect (\ref{eq:light-nu-mass}) to be applicable to the lightest neutrino mass eigenstate, and then, if the model satisfies the neutrino oscillation data global fit \cite{deSalas:2017kay}, it predicts that the neutrinoless double beta decay parameter $m_{\beta\beta}$ lies in the range
\mbox{1.2 meV $\lesssim m_{\beta\beta}\lesssim$ 3.5 meV} for the normal neutrino mass ordering and
\mbox{15 meV $\lesssim m_{\beta\beta}\lesssim$ 50  meV} for the inverted one. This is a generic result for such a small values (\ref{eq:light-nu-mass}) of the lightest neutrino state.
%In order to make the realistic model compatible with the observed neutrino mass squared differences $\delta m^2\sim10^{-5}$ and $\Delta m^2\sim10^{-3}$ we must ensure that only the mass of the lightest active neutrino is a subject of such incredibly tiny prediction at the order of magnitude $m_1\sim 10^{-13}\eV$. The other two mass eigenstates should get masses $m_{2,3}\sim 10^{-2}\eV$ in the case of inverted hierarchy, or $m_{2}\sim 10^{-3}\eV$ $m_{3}\sim 10^{-2}\eV$ in the case of normal hierarchy.
}

\subsection{Prediction for dark matter}
\label{sec:DM}
%they must be stable on cosmological time scales (otherwise they would have decayed by now), they must interact very weakly with electromagnetic radiation (otherwise they wouldn�t qualify as dark matter), and they must have the right relic density.
%
% ***** From PDG:    http://pdg.lbl.gov/2018/reviews/rpp2018-rev-dark-matter.pdf ****
%
%Candidates for non-baryonic DM in Eq. (26.1) must satisfy several conditions: they must be stable on cosmological time scales (otherwise they would have decayed by now), they must interact very weakly with electromagnetic radiation (otherwise they wouldn�t qualify as dark matter), and they must have the right relic density. Candidates include primordial black holes, axions, sterile neutrinos, and weakly interacting massive particles (WIMPs).
%
%
% ***** From our paper with Ivo:  https://arxiv.org/pdf/1712.02792.pdf
%
%the scalar field ? is the lightest Z2-odd particle. In this case, DM is produced in the early Universe via the vanilla WIMP paradigm.
%
%
\Sergey{
There is only one Dark Matter (DM) particle candidate in our model: the scalar $s^0$ specified in Eq.~(\ref{mixhHs}) as a mixture of the electroweak doublet and singlet fields.
%
%A particle candidate for dark matter should be massive and sufficiently stable and sterile. There is only one candidate in the spectrum of our model: the Higgs boson $s^0$.
}
Using the parameter fixing from the last section \ref{Pheno} we estimate its mass from \eqref{ms}, \eqref{vPhi} and \eqref{lambdaPhi}
\begin{equation}
m_s\approx 100\keV\,.
\end{equation}
\Sergey{
This is a nearly sterile state having only a tiny admixture of the doublets $H, \Sigma$ estimated as
%Its sterility is given mainly by the smallness of the admixture of the Higgs doublet fields $H$ and $\Sigma$ within $s^0$. From \eqref{UhHs} and using \eqref{kappa_est}, \eqref{lambdaFH} and \eqref{lambdaPhi}
}
\begin{eqnarray}
|\mathrm{mix}(H\in s^0)| &\approx& \frac{\lambda_{\Phi H}v_H v_\Phi-\sqrt{2}y_H M_R v_\Sigma}{\lambda_H v_{H}^2}\approx 10^{-10} \,,\\
|\mathrm{mix}(\Sigma\in s^0)| &\approx& \frac{v_\Sigma}{\sqrt{v_{\Sigma}^2+v_{\Phi}^2}} \sim \frac{1}{r_{\Phi\Sigma}} \approx 10^{-9}\,.
\end{eqnarray}
\Sergey{
In order to be a viable DM candidate its lifetime $\tau_{s}$ must be greater than
the age of the Universe $\tau_{u}$. Thus, we impose on the $s^{0}$ total decay rate $\Gamma_{s}$
the cosmological upper bound
%The stability is expressed by the decay rate $\Gamma_s$ which should be less than inverted
%age of the Universe
\begin{equation}\label{AoU}
\frac{1}{\tau_{s}} = \Gamma_s \lesssim \frac{1}{\tau_{u}}\approx 10^{-33}\eV\,.
\end{equation}
According to the recent analysis of the decaying warm DM performed in Ref.~\cite{Kuo:2018fgw}, the DM decay rate should be  smaller by other two or tree orders of magnitude over the result \eqref{AoU}, for successfully reproducing the DM abundance of the Universe of today.

For the mass value (\ref{lambdaPhi}) the only kinematically allowed decay channel is
\begin{equation}
s^0 \longrightarrow \nu\nu\,.
\end{equation}
%:{SK 19.02.2019-1} I think this part should be better eliminated from the paper.
%When $m_s>2m_e$ there is also
%\begin{equation}
%s^0 \longrightarrow e^+e^-\,,
%\end{equation}
%is allowed, but it can easily be forbidden just by making the $s^0$ a bit lighter.
%The expressions for the partial decay rates are
with the decay rate
\begin{eqnarray}
\Gamma_{s\rightarrow\nu\nu} &=& \frac{y_{s\nu\nu}^2}{8\pi}m_s \,.
%\\
%\Gamma_{s\rightarrow ee} &=& \frac{y_{see}^2}{8\pi}m_s\left(1-\frac{m_{e}^2}{m_{s}^2}\right)^2\,.
\end{eqnarray}
and with the corresponding Yukawa coupling
%Under the assumption of the hierarchy $v_\Sigma\ll v_\Phi\ll v_H$, the relevant Yukawa coupling constants are
%
\begin{eqnarray}\label{ysnunu}
y_{s\nu\nu} &=& \sqrt{2}y_\Sigma\frac{m_D}{M_R}\left(\frac{1}{r_{\Phi\Sigma}}-\frac{m_D}{M_R}\right) \approx 10^{-18} \,,
%\\
%y_{see} &=& y_H\times\mathrm{mix}(H\in s^0)\times\left(\frac{\mu_\mathrm{lin}}{M_R}-\frac{\mu_\mathrm{inv}m_D}{M_{R}^2}\right) \approx 10^{-29}\,.
\end{eqnarray}
which we evaluated, using  (\ref{mDMR}) and (\ref{rFSest}), with the result
%
%Their evaluation has been performed for the fixing of the parameters from the last subsection \ref{Pheno}. So even if the $e^+e^-$ decay channel is kinematically allowed, the dominant channel is that to neutrinos.
%
\begin{equation}
\Gamma_s \approx 10^{-32}\eV\,,
\end{equation}
assuming no significant cancelation between two terms in \eqref{ysnunu}.
%This corresponds to the lifetime one order of magnitude shorter than the age of the Universe.
%That is not so badly off for the rough order-of-magnitude estimate that we have performed here. But it is needed to say, that according to the analysis of the decaying warm dark matter performed in \cite{Kuo:2018fgw}, the dark matter decay rate should be  smaller by another two or tree orders of magnitude than \eqref{AoU} for successfully reproducing the dark matter abundance of the Universe of today.
%
%In our model, we can make the $s$-boson more stable by assuming an even more profound hierarchy than in (\ref{mDMR}), (\ref{eq:r-MR}), or by fine-tuning the free parameters, e.g. $y_H$, to cancel the dominant term in $y_{s\nu\nu}$ \eqref{ysnunu}.}
}
In our model, we can make the $s$-boson more stable by assuming an even more profound hierarchy than in (\ref{mDMR}), (\ref{eq:r-MR}), or by fine-tuning the free parameters, e.g. $y_H$, to cancel the dominant term in $y_{s\nu\nu}$ \eqref{ysnunu}, or by making the mass of the $s^0$ boson $m_s\propto v_\Phi$ smaller. The later option requires pushing $v_\Phi$ down, which eventually increases $CP$ assymmetry for the sake of leptogenesis as discussed above. We present such benchmark parameter setting in Tab.\ref{results} denoted as  {\bf CP10}.
The second option based on cancellation is demonstrated in Tab.\ref{results} as a benchmark parameter setting {\bf DMtuned1}. As one can see there, $y_{s\nu\nu}$ is two orders of magnitude lower and even with opposite sign than in the benchmark parameter setting {\bf BASIC1}, even though both benchmark settings have the same hierarchies $r_{\Phi\Sigma}$ and $m_D/M_{R}$.

\subsection{Missing energy in $Z$-boson decay}

Looking at the mass spectrum of the model, we can identify a potentially dangerous decay of the $Z$ boson
\begin{equation}
Z\longrightarrow s^0 f\bar{f} \,,
\end{equation}
since $s^0$ is not a pure electroweak singlet, but it has an admixture of the electroweak doublets \eqref{mixhHs}. Such decay process, if strong enough, would be visible at accelerators as the production of a fermion-antifermion pair plus missing energy carried away by $s^0$.

\begin{figure}[t]
	\includegraphics[width=0.9\columnwidth]{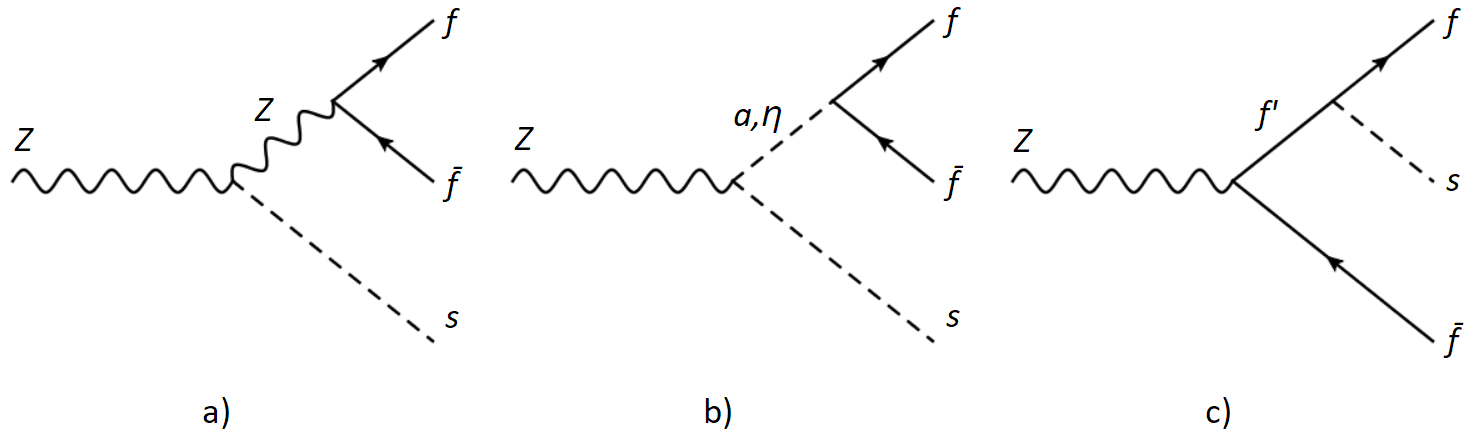}
	\caption{Feynman diagrams corresponding to the $Z$ boson decay $Z\longrightarrow s^0 f\bar{f}$: a) exchange of virtual $Z$, b) exchange of virtual $a$ and $\eta$ and c) exchange of virtual fermion $f'$.  }
	\label{missingZ}
\end{figure}

The process $Z\rightarrow s^0 f\bar{f}$ can be calculated from the tree-level amplitudes for which the exchange of virtual $Z$ boson, pseudoscalar $a$, majoron $\eta$ and fermion $f'$ should be taken into account. We show the corresponding Feynman diagrams in the Fig.~\ref{missingZ}. For charged leptons in the final state, i.e., $f=\tau,\mu,e$, the amplitude of majoron exchange vanishes as their majoron Yukawa coupling constant is $y_{\eta ff}=0$. The amplitude for the fermion exchange is completely negligible as it is proportional to the Yukawa coupling parameter $y_{sff'}\sim10^{-18}$ being estimated to be approximately same as $y_{s\nu\nu}$ given in Eq.~\eqref{ysnunu}. The contribution from the amplitude of pseudoscalar exchange is not vanishing but negligibly small together with the corresponding Yukawa coupling constant
\begin{equation}
y_{aff}=\frac{g m_f}{\sqrt{2}M_W}\frac{v_{\Sigma}v_{\Phi}}{\sqrt{v_{H}^2v_{\Sigma}^2+v_{\Sigma}^2v_{\Phi}^2+v_{\Phi}^2v_{H}^2}}\approx \mathcal{O}(10^{-13})\,.
\end{equation}
Therefore the dominant contribution comes from the $Z$ boson exchange, whose Yukawa $Zff$ coupling constant is roughly $y_{Zff}\approx\mathcal{O}(1)$, which has the most important suppression of its amplitude coming from the $H$ and $\Sigma$ mixing factors of the $s^0$ boson. Using the order-of-magnitude estimates from section \ref{Pheno}, the admixture of the SM-like Higgs doublet $H$ is at the level of $\mathcal{O}(\frac{\lambda_{\Phi H}v_\Phi}{\lambda_{H}v_H})\sim10^{-10}$ (see \eqref{epsF}, \eqref{lambdaFH} and \eqref{vPhi}), and the admixture of the doublet $\Sigma$ is at the level of $\mathcal{O}(1/r_{\Phi\Sigma})\sim10^{-9}$ \eqref{rFSest}. Assuming the $\Sigma$ admixture as the leading one, the partial decay rate for the process $Z\rightarrow s^0 f\bar{f}$ can be estimated as
\begin{equation}
\Gamma_{Z\rightarrow s^0f\bar{f}}\approx\frac{1}{r_{\Phi\Sigma}^2}10^{-3}\GeV\,.
\end{equation}
\Sergey{This gives a negligible branching ratio
\begin{equation}
B_{Z\rightarrow s^0f\bar{f}}\equiv\frac{\Gamma_{Z\longrightarrow s^0f\bar{f}}}{\Gamma_{Z}}\approx 10^{-21}\,,
\end{equation}
compatible with the experimental data  \cite{PhysRevD.98.030001}. Here $\Gamma_{Z}\doteq2.5\GeV$ is the total decay width of the $Z$ boson.
%The result is completely negligibly sensitive to the lepton mass.

For the same reason, namely because of smallness of the electroweak non-singlet component of the $s^0$-boson, other invisible $Z$ boson decay channels like, e.g., $Z\rightarrow s^0s^0 f\bar{f}$, are also totally negligible.
}

%For the same reason, namely because of the sterility of the $s^0$ boson, also other invisible $Z$ boson decay channels like, e.g., $Z\rightarrow s^0s^0 f\bar{f}$, are totally negligible as well.

\section{Conclusions}

In this work, the possibility of dynamical LNV and neutrino mass generation, based on neutrino condensation, has been considered.
In order to proof this concept, a simplified
%beyond-SM
model setup with a single neutrino generation, to wit, without neutrino flavor mixing, has been
%proposed and
constructed and studied.
%, although for the sake of simplicity the model contains only a single neutrino generation, and therefore it ignores the phenomenon of neutrino mixing.
This test setup also lacks $CP$ violation in the neutrino Yukawa sector, and then it invalidates itself from being able to describe leptogenesis. Nevertheless, we believe that it shares important qualitative features and the order-of-magnitude quantitative estimates with a realistic three-generation version of the model, which is going to be developed in a successive work.

In order to check the viability of our neutrino condensation model, we have borrowed realistic leptogenesis constraints on the size of the mass of quasi-degenerate heavy Majorana neutrinos, their mass splitting and decay rates.
We assumed a combined inverse plus linear seesaw mechanism for the explanation of the active neutrino mass and kept the seesaw scale rather low, i.e., $M_R\sim1-10\TeV$, in order stay in the ballpark of current and near-future collider experiments.
%
%Since we want to be in the ballpark of current and near-future colliders data, we assumed a combined inverse plus linear seesaw mechanism for the explanation of the active neutrino mass and keep the seesaw scale rather low, i.e., $M_R\sim1-10\TeV$.
The neutrino condensation dynamically generates the LNV mass entries of the seesaw mass matrix, as VEVs of composite additional Higgs fields. Their coupling parameters are generated dynamically and fixed completely by the underlying new dynamics, which provides the necessary attraction within the LNV neutrino-neutrino channels.

The main message of this work, based on order-of-magnitude estimates, is that, in spite of such tightly constrained scheme with limited
%no
room to play with its parameters, the model has the potential to predict viable values of active neutrino masses and the mass and decay rate of a dark matter particle candidate, while satisfying the parameter requirements of successful leptogenesis.

\vskip5mm

\centerline{\bf Acknowledgements}

\medskip

This work was supported by Fondecyt (Chile) under grants No. 1170171, No. 3150472,
No. 1180232, No. 1190845 as well as \mbox{CONICYT} (Chile)  Basal FB0821.
The work was supported from European Regional Development Fund-Project Engineering Applications of Microworld Physics (No. CZ.02.1.01/0.0/0.0/16\_019/0000766).

\appendix

\section{Neutrino mass matrix diagonalization}\label{AppM}

We present here the eigenvalues of the full neutrino mass matrix in Eq.~\eqref{M}, i.e., including the non-zero $\mu'_\mathrm{inv}$-entry, under the assumption of the hierarchy \eqref{seesaw_hierarchy}, which comes from our phenomenological analysis done in Sect.~\ref{Pheno}. Here, we add the assumption
\begin{equation}
\mu'_\mathrm{inv}\sim \mu_\mathrm{inv} \,.
\end{equation}
We perform an expansion of the mass eigenvalues in the following small parameters, dubbed as $\varepsilon$:
\begin{equation}\label{epsilon_hierarchy}
\varepsilon
\sim\sqrt{\frac{\mu_\mathrm{lin}}{M_R}}
\sim\frac{m_{D}}{M_{R}}
\sim\frac{\mu_{\mathrm{inv}}}{M_{R}}
\sim\frac{{\mu'_{\mathrm{inv}}}}{M_{R}}
% \sim\frac{\mu_{\mathrm{lin}}}{m_{D}}
% \sim\frac{\mu_{\mathrm{lin}}}{\mu_{\mathrm{inv}}}
% \sim\frac{\mu_{\mathrm{lin}}}{{\mu'_{\mathrm{inv}}}}
\,,
\end{equation}
To leading order in all of these small parameters the eigenvalues are
\begin{subequations}\label{mnu_App}
\begin{eqnarray}
\frac{m_{\nu}}{M_R} &\simeq& \left(\frac{m_D}{M_R}-\frac{m_D^3}{M_{R}^3}\right)\left(\frac{\mu_{\rm inv}}{M_R} \frac{m_D}{M_{R}}-2\frac{\mu_{\rm lin}}{M_R}\right) \nonumber\\
&&\quad\quad+
\left(\frac{\mu_{\mathrm{lin}}^2}{M_{R}^2}-2\frac{\mu_\mathrm{lin}}{M_R}\frac{m_{D}}{M_{R}}\frac{\mu_{\mathrm{inv}}}{M_{R}}+\frac{m_{D}^2}{M_{R}^2}\frac{\mu_{\mathrm{inv}}^2}{M_{R}^2}\right)\frac{\mu'_{\mathrm{inv}}}{M_R}+\mathcal{O}(\varepsilon^5) \,,\label{mnux} \nonumber\\
\frac{m_{N_\pm}}{M_R} &\simeq&  \pm 1+\frac{1}{2}\frac{(\mu_\mathrm{inv}+\mu'_{\mathrm{inv}})}{M_{R}}\pm\frac{4m_{D}^2+(\mu_\mathrm{inv}-\mu'_{\mathrm{inv}})^2}{8M_{R}^2} \nonumber\\
&&\quad\quad-\frac{1}{2}\frac{m_D}{M_R}\left(\frac{\mu_{\rm inv}}{M_R} \frac{m_D}{M_{R}}-2\frac{\mu_{\rm lin}}{M_R}\right)+\mathcal{O}(\varepsilon^3) \,.\nonumber
\end{eqnarray}
\end{subequations}
The light neutrino mass, to leading order $\mathcal{O}(\varepsilon^3)$ and heavy neutrino masses to order $\mathcal{O}(\varepsilon^1)$, are given as
\begin{eqnarray}
%m_{\nu} &\simeq& (\nu m_D-2\mu M_R)\frac{m_D}{M_{R}^2} \,,
\frac{m_{\nu}}{M_R} &\simeq& \frac{\mu_{\rm inv}}{M_{R}}\frac{m_{D}^2}{M_{R}^2}-2\frac{\mu_{\rm lin}}{M_R}\frac{m_{D}}{M_{R}} \,,\label{mnu3}\\
\frac{m_{N_\pm}}{M_R} &\simeq&  \pm 1+\frac{1}{2}\frac{\mu_\mathrm{inv}+\mu'_{\mathrm{inv}}}{M_{R}} \,. \label{mN1}
\end{eqnarray}
The expression \eqref{mnu3} for light neutrino mass coincides with Eq.~\eqref{mnu} and it does not depend on $\mu'_{\mathrm{inv}}$, in accordance with \cite{Gavela:2009cd,Blanchet:2009kk,Blanchet:2012bk}. That motivates our assumption of $\mu'_{\mathrm{inv}}=0$, under which the expression \eqref{mN1} of heavy neutrino mass coincides with Eq.~\eqref{mnu}.

The neutrino mass eigenstates $(\nu, N_{-}, N_{+})$ are linear combinations of the original fields
\begin{eqnarray}\label{mixnu}
\beginm{c}\nu \\ N_- \\ N_+ \endm &\simeq&
\mathcal{U}_\nu \beginm{c} \nu_L \\ \nu_{R}^\C \\ S_{R}^\C \endm  \,,
\end{eqnarray}
where $\mathcal{U}_\nu$ is the neutrino mixing matrix transforming the neutrino mass matrix \eqref{M} into its diagonal form $\mathcal{U}_\nu M_\nu \mathcal{U}_{\nu}^\T$. To lowest order of the $\varepsilon$-parameters the neutrino mixing matrix is
\begin{eqnarray}\label{Unu}
\mathcal{U}_\nu &\simeq&
\beginm{ccc} -1+\frac{m_{D}^2}{2M_{R}^2} & \frac{\mu_\mathrm{lin}}{M_R}-\frac{\mu_\mathrm{inv}m_D}{M_{R}^2} & \frac{m_{D}}{M_{R}} \\
\frac{m_{D}}{\sqrt{2}M_{R}} & -\frac{1}{\sqrt{2}} & \frac{1}{\sqrt{2}}\left(1-\frac{m_{D}^2}{2M_{R}^2}\right) \\
\frac{m_{D}}{\sqrt{2}M_{R}} &  \frac{1}{\sqrt{2}} & \frac{1}{\sqrt{2}}\left(1-\frac{m_{D}^2}{2M_{R}^2}\right) \endm \,.
\end{eqnarray}

\section{Higgs boson mass matrices}\label{App_Higgs}

The mass matrices of the Higgs bosons are obtained by Eq.~\eqref{M_Higgs} from the effective potential \eqref{V}.
\begin{itemize}

\item The $(2\times2)$ mass matrix of charged Higgs bosons is obtained from
\begin{equation}
\Big[M_{\mathrm{charged}}^2\Big]_{ij}=\left.\frac{\partial^2}{\partial a_{i}^-\partial a_{j}^+} \V_\mathrm{eff}(0,0,a_{H}^-,a_{H}^+,0,0,a_{\Sigma}^-,a_{\Sigma}^+,0,0) \right|_{a_{H}^\pm=0,a_{\Sigma}^\pm=0}\,,\ \mathrm{for\ }i,j=H,\Sigma\,.
\end{equation}
The resulting mass matrix is
\begin{equation}\label{M_charged}
M_{\mathrm{charged}}^2 = \frac{1}{2}(\sqrt{2}v_\Phi\kappa-\lambda'_{H\Sigma}v_H v_\Sigma)\beginm{cc} \frac{v_\Sigma}{v_H} & 1 \\ 1 & \frac{v_H}{v_\Sigma} \endm \,,
\end{equation}
which is transformed into the diagonal form $\mathcal{U}_\mathrm{charged}M_{\mathrm{charged}}^2\mathcal{U}_\mathrm{charged}^\T$ by means of the orthogonal matrix
\begin{eqnarray} \label{Ucharged}
\mathcal{U}_\mathrm{charged} =
\frac{1}{v}\beginm{cc} -v_H & v_\Sigma  \\
v_\Sigma & v_H \endm \,.
\end{eqnarray}

\item The $(3\times3)$ mass matrix neutral pseudo-scalar Higgs bosons is obtained from
\begin{equation}
\Big[M_{\mathrm{pseudo}}^2\Big]_{ij}=\left.\frac{\partial^2}{\partial a_{i}^0\partial a_{j}^0} \V_\mathrm{eff}(0,a_{H}^0,0,0,0,a_{\Sigma}^0,0,0,0,a_{\Phi}^0) \right|_{a_{H}^0=0,a_{\Sigma}^0=0,a_{\Phi}^0=0}\,,\ \mathrm{for\ }i,j=H,\Sigma,\Phi\,.
\end{equation}
The resulting mass matrix is then
\begin{equation}\label{M_pseudo}
M_{\mathrm{pseudo}}^2 = \frac{\kappa}{\sqrt{2}}\beginm{ccc} \frac{v_\Sigma v_\Phi}{v_H} & v_\Phi & v_\Sigma \\ v_\Phi & \frac{v_H v_\Phi}{v_\Sigma} & v_H \\ v_\Sigma & v_H & \frac{v_H v_\Sigma}{v_\Phi} \endm \,,
\end{equation}
which is transformed into the diagonal form $\mathcal{U}_\mathrm{pseudo}M_{\mathrm{pseudo}}^2\mathcal{U}_\mathrm{pseudo}^\T$ by means of the orthogonal matrix
\begin{eqnarray} \label{Upseudo}
\mathcal{U}_\mathrm{pseudo} =
\beginm{ccc} \frac{v_H}{v} & -\frac{v_\Sigma}{v} & 0 \\
0 & -\frac{v_\Sigma}{\sqrt{v_{\Sigma}^2+v_{\Phi}^2}} & \frac{v_\Phi}{\sqrt{v_{\Sigma}^2+v_{\Phi}^2}} \\
\frac{v_\Sigma v_{\Phi}}{\sqrt{v_{H}^2v_{\Sigma}^2+v_{\Sigma}^2v_{\Phi}^2+v_{\Phi}^2v_{H}^2}} &
\frac{v_{\Phi}v_H}{\sqrt{v_{H}^2v_{\Sigma}^2+v_{\Sigma}^2v_{\Phi}^2+v_{\Phi}^2v_{H}^2}} &
\frac{v_H v_\Sigma}{\sqrt{v_{H}^2v_{\Sigma}^2+v_{\Sigma}^2v_{\Phi}^2+v_{\Phi}^2v_{H}^2}} \endm \,.
\end{eqnarray}

\item The $(3\times3)$ mass matrix neutral scalar Higgs bosons is obtained from
\begin{equation}
\Big[M_{hHs}^2\Big]_{ij}=\left.\frac{\partial^2 }{\partial h_{i}^0\partial h_{j}^0}\V_\mathrm{eff}(h_{H}^0,0,0,0,h_{\Sigma}^0,0,0,0,h_{\Phi}^0,0)\right|_{h_{H}^0=0,h_{\Sigma}^0=0,h_{\Phi}^0=0}\,,\ \mathrm{for\ }i,j=H,\Sigma,\Phi\,.
\end{equation}
The resulting mass matrix is then
\begin{equation}\label{M_hHs}
M_{hHs}^2 = \beginm{ccc}
\frac{v_\Sigma v_\Phi \kappa}{\sqrt{2}v_H}+v_{H}^2\lambda_H &
-\frac{v_\Phi \kappa}{\sqrt{2}}+v_{H}v_\Sigma(\lambda_{H\Sigma}+\lambda'_{H\Sigma}) &
-\frac{v_\Sigma \kappa}{\sqrt{2}}+v_{H}v_\Phi\lambda_{\Phi H} \\
-\frac{v_\Phi \kappa}{\sqrt{2}}+v_{H}v_\Sigma(\lambda_{H\Sigma}+\lambda'_{H\Sigma}) &
\frac{v_H v_\Phi \kappa}{\sqrt{2}v_\Sigma}+v_{\Sigma}^2\lambda_\Sigma &
-\frac{v_H \kappa}{\sqrt{2}}+v_{\Sigma}v_\Phi\lambda_{\Phi \Sigma} \\
-\frac{v_\Sigma \kappa}{\sqrt{2}}+v_{H}v_\Phi\lambda_{\Phi H} &
-\frac{v_H \kappa}{\sqrt{2}}+v_{\Sigma}v_\Phi\lambda_{\Phi \Sigma} &
\frac{v_H v_\Sigma \kappa}{\sqrt{2}v_\Phi}+v_{\Phi}^2\lambda_\Phi \endm \,,
\end{equation}
which is transformed into the diagonal form $\mathcal{U}_{hHs}M_{hHs}^2\mathcal{U}_{hHs}^\T$ by means of the orthogonal matrix which, under the hierarchy $v_\Sigma\ll v_\Phi\ll v_H$ ($r_{\Phi\Sigma}\gg1$) and $\lambda_{\Phi H}\ll\lambda_H$, is approximated as
\begin{eqnarray} \label{UhHs}
\mathcal{U}_{hHs} =
\beginm{ccc}
1-\varepsilon_H & \varepsilon_\Sigma & \varepsilon_\Phi \\
\varepsilon_\Sigma & -\frac{v_\Phi}{\sqrt{v_{\Sigma}^2+v_{\Phi}^2}} & \frac{v_\Sigma}{\sqrt{v_{\Sigma}^2+v_{\Phi}^2}} \\
-\varepsilon_\Phi & \frac{v_\Sigma}{\sqrt{v_{\Sigma}^2+v_{\Phi}^2}} & \frac{v_\Phi}{\sqrt{v_{\Sigma}^2+v_{\Phi}^2}}
\endm \,,
\end{eqnarray}
where
\begin{eqnarray}
\varepsilon_H &=&  \frac{1}{2}\left(\frac{\big(\lambda_{\Phi H}v_H v_\Phi-\sqrt{2}\kappa v_\Sigma\big)^2}{\lambda_{H}^2 v_{H}^4}-\frac{v_{\Sigma}^2}{v_{H}^2}\right)\,, \\
\varepsilon_\Sigma &=& \frac{v_\Sigma}{v_H} \,, \\
\varepsilon_\Phi &=& \frac{\lambda_{\Phi H}v_H v_\Phi-\sqrt{2}\kappa v_\Sigma}{\lambda_H v_{H}^2} \,.\label{epsF}
\end{eqnarray}

\end{itemize}

\section{Renormalization Group Equations}
\label{AppRGE}
\Sergey{
We derived the Renormalization Group Equations (RGE), used in our analysis, with the help of the pyR@TE software \cite{Lyonnet:2013dna,Lyonnet:2016xiz}.
%
%By means of the pyR@TE software \cite{Lyonnet:2013dna,Lyonnet:2016xiz} we have obtained the RG equations.
%
}

The RGEs for the gauge coupling constants of hypercharge $g_1$, of $\SU{2}_L$ $g_2$, and of color $g_3$, are the same as for the two-Higgs-doublet models,
\begin{subequations}\label{g123}
	\begin{eqnarray}
	\D g_1 = \hphantom{-}7g_{1}^3 \,,&\ \ &g_{1}^2(M_Z)\doteq0.127 \,, \\
	\D g_2 = -3g_{2}^3 \,,&\ \ &g_{2}^2(M_Z)\doteq0.425 \,, \\
	\D g_3 = -7g_{3}^3 \,,&\ \ &g_{3}^2(M_Z)\doteq1.440 \,,
	\end{eqnarray}
\end{subequations}
where
\begin{equation}
\D\equiv 16\pi^2\frac{\d}{\d t} \,
\end{equation}
and $t$ is
\begin{equation}
t=\ln\frac{m}{M_Z}\,.
\end{equation}
In the RGE evolution we neglect the effect of the Yukawa coupling constants other than the neutrino Yukawa couplings $y_H$, $y_\Sigma$ and $y_\Phi$ from Eq.~\eqref{L} and the SM Yukawa coupling of the top quark, $y_t$. The latter runs from the compositeness scale $\Lambda$ all the way down to the electroweak scale, according to the following RGE:
\begin{equation}
\D y_t    = y_t\big[\tfrac{9}{2}y_{t}^2+\theta(t-t_{M_R})y_{H}^2-\tfrac{17}{12}g_{1}^2-\tfrac{9}{4}g_{2}^2-8g_{3}^2\big] \,, \label{yt_RGE} \\
\end{equation}
where
\begin{equation}
t_{M}=\ln\frac{M}{M_Z}\,.
\end{equation}
%{\color{red} What is $t_{M_R}$?}
On the other hand, the neutrino Yukawa couplings run according to their RGEs
\begin{eqnarray}
\D y_H & = & y_H\big[\tfrac{1}{2}\big(5y_{H}^2+y_{\Sigma}^2\big)+3y_{t}^2-\tfrac{3}{4}g_{1}^2-\tfrac{9}{4}g_{2}^2\big] \,,  \\
\D y_\Sigma & = & y_\Sigma\big[\tfrac{1}{2}\big(5y_{\Sigma}^2+y_{H}^2+4y_{\Phi}^2\big)-\tfrac{3}{4}g_{1}^2-\tfrac{9}{4}g_{2}^2\big] \,,  \\
\D y_\Phi & = & y_{\Phi}\big(6y_{\Phi}^2+2y_{\Sigma}^2\big) \,,
\end{eqnarray}
only down to the right-handed neutrino mass scale $M_R$, where the heavy neutrinos decouple. At that scale we trade the neutrino Yukawa coupling constants for the effective generalized non-renormalizable Weinberg operators. Here we write the RGEs only for the two operators which are relevant
and give a leading order contribution to the active neutrino masses. They are
\begin{eqnarray}
\D w_{H\Phi H} & = & w_{H\Phi H}\big[6\, \theta(t-t_{m_t})y_{t}^2 + \lambda_H +2 \lambda_{H\Phi} -3\, \theta(t-t_{M_Z})g_{2}^2\big] \,,  \\
\D w_{H\Sigma} & = & w_{H\Sigma}\big[3\, \theta(t-t_{m_t})y_{t}^2 +\lambda_{H\Sigma} +\lambda'_{H\Sigma} -3\, \theta(t-t_{M_Z})g_{2}^2\big] \,.
\end{eqnarray}
Here we have introduced thresholds corresponding to $m_t$ and $M_Z$, because in order to determine the neutrino masses, we need to run the Weinberg parameters many orders of magnitude below the electroweak scale. Keeping the coupling constants $y_t$ and $g_2$ would affect the running of the Weinberg parameters significantly and unphysically.

The RGEs for the dimensionless couplings $\lambda$'s of the effective potential in Eq.~\eqref{V} are
\begin{eqnarray}
\D\lambda_H & = & 12\lambda_{H}^2 + 4\lambda_{H\Sigma}^2 + 4\lambda_{H\Sigma}\lambda'_{H\Sigma} + 2{\lambda'_{H\Sigma}}^2 + 2\lambda_{\Phi H}^2 + \nonumber\\
&& - 3\lambda_{H}(3g_{2}^2+g_{1}^2)+\frac{3}{2}g_{2}^4+\frac{3}{4}(g_{2}^2+g_{1}^2)^2 + \nonumber\\
&& + 4\lambda_{H}\big[\theta(t-t_{M_R})y_{H}^2+3y_{t}^2\big]-4\, \theta(t-t_{M_R})y_{H}^4-12y_{t}^4 \,,\\
\D\lambda_\Sigma & = & 12\lambda_{\Sigma}^2 + 4\lambda_{H\Sigma}^2 + 4\lambda_{H\Sigma}\lambda'_{H\Sigma} + 2{\lambda'_{H\Sigma}}^2 + 2\lambda_{\Phi \Sigma}^2 + \nonumber\\
&&  - 3\lambda_{\Sigma}(3g_{2}^2+g_{1}^2)+\frac{3}{2}g_{2}^4+\frac{3}{4}(g_{2}^2+g_{1}^2)^2 + \nonumber\\
&& +4\lambda_\Sigma \theta(t-t_{M_R})y_{\Sigma}^2- 4\, \theta(t-t_{M_R})y_{\Sigma}^4 \,,\\
\D\lambda_{H\Sigma} & = & 6\lambda_{H}\lambda_{H\Sigma} +2\lambda_{H}\lambda'_{H\Sigma}+ 6\lambda_{H\Sigma}\lambda_{\Sigma}+ 2\lambda_{\Sigma}\lambda'_{H\Sigma} + 4\lambda_{H\Sigma}^2+2{\lambda'_{H\Sigma}}^2 + 2\lambda_{\Phi H}\lambda_{\Phi \Sigma} + \nonumber\\
&& - 3\lambda_{H\Sigma}(3g_{2}^2+g_{1}^2)+\frac{9}{4}g_{2}^4+\frac{3}{4}g_{1}^4-\frac{3}{2}g_{2}^2g_{1}^2 + \nonumber\\
&& +2\lambda_{H\Sigma}\big[\theta(t-t_{M_R})\big(y_{H}^2+y_{\Sigma}^2\big)+3y_{t}^2\big] \,,\\
\D\lambda'_{H\Sigma} & = & 2\lambda_{H}\lambda'_{H\Sigma} +2\lambda_{\Sigma}\lambda'_{H\Sigma}+8\lambda_{H\Sigma}\lambda'_{H\Sigma} + 4{\lambda'_{H\Sigma}}^2 +  \nonumber\\
&&  - 3\lambda'_{H\Sigma}(3g_{2}^2+g_{1}^2)+3g_{2}^2g_{1}^2 + \nonumber\\
&& +2\lambda'_{H\Sigma}\big[\theta(t-t_{M_R})\big(y_{\Sigma}^2+y_{H}^2)+3y_{t}^2\big]-\tfrac{7}{2}\, \theta(t-t_{M_R})y_{H}^2y_{\Sigma}^2 \,,\\
\D\lambda_{\Phi} & = & 10\lambda_{\Phi}^2 + 4\lambda_{\Phi H}^2 + 4\lambda_{\Phi \Sigma}^2  +8\, \theta(t-t_{M_R})\lambda_\Phi y_{\Phi}^2 -32\, \theta(t-t_{M_R})y_{\Phi}^4 \,,\\
\D\lambda_{\Phi H} & = & 4\lambda_{\Phi H}^2 + 6\lambda_{H}\lambda_{\Phi H} + 4\lambda_{\Phi}\lambda_{\Phi H} + 4\lambda_{\Phi \Sigma}\lambda_{H \Sigma} + 2\lambda_{\Phi\Sigma}\lambda'_{H\Sigma} + \nonumber\\
&&   - \frac{3}{2}\lambda_{\Phi H}(3g_{2}^2+g_{1}^2) + 2\lambda_{\Phi H}\big[\theta(t-t_{M_R})\big(y_{H}^2+2y_{\Phi}^2\big)+3y_{t}^2\big] \,,\\
\D\lambda_{\Phi \Sigma} & = & 4\lambda_{\Phi \Sigma}^2 + 6\lambda_{\Sigma}\lambda_{\Phi \Sigma} + 4\lambda_{\Phi}\lambda_{\Phi \Sigma} + 4\lambda_{\Phi H}\lambda_{H \Sigma} + 2\lambda_{\Phi H}\lambda'_{H\Sigma} + \nonumber\\
&&   - \frac{3}{2}\lambda_{\Phi \Sigma}(3g_{2}^2+g_{1}^2) +2\lambda_{\Phi\Sigma}\theta(t-t_{M_R})\big(y_{\Sigma}^2+2y_{\Phi}^2\big)-16\, \theta(t-t_{M_R})\, y_{\Phi}^2\, y_{\Sigma}^2 \,.
\end{eqnarray}
Finally, the RGE for the dimensionfull coupling parameter $\kappa$ of the effective potential \eqref{V} is
\begin{eqnarray}\label{RGEkappa}
\D\kappa & = & \kappa\big[2\lambda_{\Phi H} + 2\lambda_{\Phi\Sigma} + 2\lambda_{H\Sigma} + 4\lambda'_{H\Sigma} - \tfrac{3}{2}\big(3g_{2}^{2}+g_{1}^{2}\big) \\
&& \ \ \  + \theta(t-t_{M_R})\big(y_{H}^2 + y_{\Sigma}^2 + 2y_{\Phi}^2\big) + 3 y_{t}^2\big] - 8 \, \theta(t-t_{M_R})\, y_{H}\, y_{\Sigma}\, y_{\Phi}\, M_R \nonumber
\end{eqnarray}
In the RGEs for the couplings  $\lambda$'s and $\kappa$ we introduce just one threshold, corresponding to $M_R$. The thresholds around the electroweak scale do not play a significant role in determining the mass spectrum for the Higgs bosons, the top-quark and the electroweak gauge bosons, as they all lie in the same ballpark.

%{\color{red} (In the $\theta$-functions, where it says $M_R$, etc... , e.g. $\theta(t-M_R)$, it should say $t_{M_R}$...right?)}

\section{Numerical solutions of the RGEs }
\label{example}

We show here one of the viable examples of numerical solution of the model.

\noindent Input RGE boundary conditions:
\begin{eqnarray}
\Lambda &=& 100\TeV \\
y_{\Sigma,\Phi}(\Lambda) &=& 3\\
\lambda_{\Sigma,\Phi,H\Sigma,H\Sigma',\Phi H,\Phi\Sigma}(\Lambda)&=&0 \\
\kappa(\Lambda)&=&0
\end{eqnarray}
\noindent Input SM parameters:
\begin{eqnarray}
v &= & 246\GeV\ \ \longrightarrow\ \ v_H\doteq 246\GeV \\
m_{h^0} & = & 125\GeV\ \ \longrightarrow\ \ \lambda_H(m_h) = 0.258 \\
m_t &= & 174\GeV\ \ \longrightarrow\ \ y_t(m_t)\doteq 1.0003
\end{eqnarray}
In Table~\ref{results} we present four benchmark parameter settings. The benchmark parameter setting {\bf BASIC10} corresponds to our order-of-magnitude estuimate performed in the section Sec.~\ref{Pheno}. The {\bf BASIC1} setting is included in order to show the impact of decreasing the value of $M_R$. The {\bf DMtuned1} setting is shown in order to demonstrate the cancellation in the DM decay Yukawa coupling constant $y_{s\nu\nu}$ from Eq.~\eqref{ysnunu}. The {\bf CP10} setting is included in order to show the impact of requirement to increase the $CP$ assymmetry up to the level $\sim0.01$.

\begin{table}[h]
	\begin{center}
		\begin{tabular}{|l|c|c|c|c|}
			\hline
			\hline
			\begin{tabular}{c}{\bf benchmark} \\ {\bf param. sets}\end{tabular} & {\bf BASIC10} & {\bf BASIC1} & {\bf DMtuned1} & {\bf CP10} \\
			\hline
			\hline
			$\ M_R[\TeV]$ & $10$ & $1.11$ & $1.11$ & $10$ \\
			$\ v_\Phi[\keV]$ & $120$ & $120$ & $0.1$ & $0.012$ \\
			$\ r_{\Phi\Sigma}\ ||\ v_\Sigma[\meV]$ & $1.2\times 10^9\ ||\ 0.1$ & $1.2\times 10^8\ ||\ 1$ & $10^8\ ||\ 0.001$ & $1.2\times 10^9\ ||\ 0.00001$ \\
			$\ y_H(M_R)\ ||\ m_D[\keV]$ & $10^{-7}\ ||\ 17.4$ & $10^{-7}\ ||\ 17.4$  & $0.9\times10^{-7}\ ||\ 15.7$ & $10^{-7}\ ||\ 17.4$ \\
			\hline
			$\ y_\Sigma(M_R)\ ||\ \mu_\mathrm{lin}[\meV]$ & $2.12\ ||\ 0.15$ & $1.78\ ||\ 1.26$ & $1.78\ ||\ 0.0013$ & $2.12\ ||\ 0.000015$ \\
			$\ y_\Phi(M_R)\ ||\ \mu_\mathrm{inv}[\keV]$ & $0.17\ ||\ 1404$ & $1.26\ ||\ 107$ & $1.26\ ||\ 0.09$ & $1.65\ ||\ 0.014$ \\
			$\ m_{\nu}^{(\mathrm{lin})}[\eV]$ & $1.03\times 10^{-12}$ & $0.08\times 10^{-9}$ & $7.1\times 10^{-14}$ & $1.03\times 10^{-16}$ \\
			$\ m_{\nu}^{(\mathrm{inv})}[\eV]$ & $0.74\times 10^{-12}$ & $0.05\times 10^{-9}$ & $3.3\times 10^{-14}$ & $0.74\times 10^{-16}$ \\
			$\ m_\nu[\eV]$ & $0.29\times 10^{-12}$ & $0.03\times 10^{-9}$ & $3.9\times 10^{-14}$ & $0.29\times 10^{-16}$ \\
			\hline
			$\ \kappa(M_Z)[\MeV]$ & $0.57$ & $0.09$ & $0.083$ & $0.57$ \\
			\hline
			$\ \lambda_H(\Lambda)$ & $0.0068$ & $0.0068$ & $0.0068$ & $0.0068$ \\
			$\ \lambda_\Sigma(M_Z)$ & $1.151$ & $1.356$ & $1.356$ & $1.151$ \\
			$\ \lambda_\Phi(M_Z)$ & $2.044$ & $2.265$ & $2.265$ & $2.044$ \\
			$\ \lambda_{\Phi\Sigma}(M_Z)$ & $0.152$ & $0.171$ & $0.171$ & $0.152$ \\
			$\ \lambda_{H\Sigma}(M_Z)$ & $-0.010$ & $-0.009$ & $-0.009$ & $-0.010$ \\
			$\ \lambda'_{H\Sigma}(M_Z)$ & $-0.006$ & $-0.006$ & $-0.006$ & $-0.006$ \\
			$\ \lambda_{\Phi H}(M_Z)$ & $0.0002$ & $0.0002$ & $0.0002$ & $0.0002$ \\
			\hline
			$\ m_{H^0,a^0,h^\pm}[\TeV]$ & $10.87$ & $1.37$ & $1.198$ & $10.87$ \\
			$\ (m_{h^\pm}-m_{H^0})[\MeV]$ & $2.5$ & $38.4$ & $45.4$ & $2.5$ \\
			$\ (m_{H}-m_{a^0})[\eV]$ & $0.0$ & $0.0$ & $0.0002$ & $0.0018$ \\
			\hline
			$\ y_{N\nu h}$ & $5.0\times 10^{-8}$ & $5.0\times 10^{-8}$ & $4.5\times 10^{-8}$ & $5.0\times 10^{-8}$ \\
			$\ y_{N\nu s}$ & $2.0\times 10^{-9}$ & $1.2\times 10^{-8}$ & $0.9\times 10^{-8}$ & $2.0\times 10^{-9}$ \\
			$\ y_{N\nu \eta}$ & $2.0\times 10^{-9}$ & $1.2\times 10^{-8}$ & $0.9\times 10^{-8}$ & $2.0\times 10^{-9}$ \\
			\hline
			$\ m_{s^0}[\keV]$ & $103$ & $105$ & $0.075$ & $0.0089$ \\
			$\ y_{s\nu\nu}$ & $-2.74\times 10^{-18}$ & $-1.08\times 10^{-16}$ & $1.43\times 10^{-18}$ & $-2.74\times 10^{-18}$ \\
			$\ \Gamma_s[\eV]$ & $3.06\times10^{-32}$ & $4.86\times10^{-29}$ & $6.12\times10^{-36}$ & $2.56\times10^{-36}$ \\
			\hline
			\hline
		\end{tabular}
	\end{center}
	\caption[]{\small Table of benchmark parameter settings. }
	\label{results}
\end{table}

\bibliography{Ref_Neutrino_condensation}

\end{document}